# An Empirical Correction for Moderate Multiple Scattering in Super-Heterodyne Light Scattering

Denis Botin[1,*], Ludmila Marotta[1,2], Holger Schweinfurth[1], Bastian Sieber[1], Christopher Wittenberg[1] and Thomas Palberg[1]

[1]*Institut für Physik, Johannes Gutenberg Universität, D-55099 Mainz, Germany*
[2]*Institute of Chemical Engineering, Federal University of Itajubá*

**Frequency domain super-heterodyne laser light scattering is utilized in a low angle integral measurement configuration to determine flow and diffusion in charged sphere suspensions showing moderate to strong multiple scattering. We introduce an empirical correction to subtract the multiple scattering background and isolate the singly scattered light. We demonstrate the excellent feasibility of this simple approach for turbid suspensions of transmittance T ≥ 0.4. We study the particle concentration dependence of the electro-kinetic mobility in low salt aqueous suspension over an extended concentration regime and observe a maximum at intermediate concentrations. We further use our scheme for measurements of the self-diffusion coefficients in the fluid samples in the absence or presence of shear, as well as in polycrystalline samples during crystallization and coarsening. We discuss the scope and limits of our approach as well as possible future applications.**

**PACS: if needed**

**Corresponding author: Denis Botin dbotin@uni-mainz.de**

# INTRODUCTION

Multiple scattering (MS) strongly affects studies of turbid colloidal suspensions using Laser light scattering. Depending on the degree of MS, several different sophisticated approaches have been taken to use, correct for, or suppress MS in studies on suspension dynamics. At very large turbidities and small optical path lengths, suspension dynamics can be determined using diffusive wave spectroscopy (DWS) [1[2]]. In the regime of low to moderate multiple scattering, index matching is regularly employed [3, 4, 5], but it is not easily applicable in water based systems. Further, optical path lengths may be shortened by utilizing fibre optics [6]. In addition, cross correlation schemes [7] were pioneered by Phillies in the early eighties [8] and further developed to two-color or 2D cross correlation schemes [9, 10]. This way the methods and theory developed for (single) dynamic light scattering in the time domain [11] could be used also in turbid samples. Alternatively, in frequency domain, special mode selective heterodyne instrumentation was developed and applied to Brilluoin scattering [12, 13, 14]. Both cross correlation and mode selection, however, afford complex instrumentation and data evaluation schemes. A much simpler instrumentation is needed for the statistical analysis of heterodyne speckle fields taken at different times providing information about the velocity field in the fluid in a given plane perpendicular to the optical axis [15]. This information can be extracted, either by measuring their cross-correlation function or by recovering the power spectrum corresponding to the difference between the two speckle fields. Multiple scattering in this approach is minimized by using confocal geometry [16]. Yet another simple approach is path length resolved low coherence interferometry, which also is a static heterodyne technique [17, 18]. There a certain small path length can be selected which corresponds to a single back-scattering event. Then only singly scattered light is recorded and can be analyzed for sample dynamics. For larger path lengths and weakly forward scattering samples, however, the scattering vector becomes ill defined due to the detection of additional multiply scattered photons. In the present paper we are concerned with Laser Doppler velocimetry, a far field heterodyne frequency domain technique to study suspension flow. This technique will be used in a standard super-heterodyne low angle version [19] with spatially restricted detection volume for electro-kinetic studies over an extended range of volume fractions. We study samples of very different scattering contrast and turbidity. To prepare the ground for future systematic studies, we have been particularly interested in minimizing any extra instrumental complexity.

Doppler velocimetry is a well-established frequency domain technique used to study the flow properties of colloidal dispersions [11]. A typical application is the determination of the electro-phoretic mobility, $\mu_{ep}$, from the particle drift velocity, $\mathbf{v}_{ep} = \mu_{ep} \cdot \mathbf{E}$, of a particle in the applied electric field $\mathbf{E}$. The mobility can be interpreted in terms of the particle zeta potential and effective charge [20, 21]. Previous research has revealed two differing types of particle concentration dependence for this central electro-kinetic quantity. In studies on isolated particles in either aqueous [22] or organic solvent [23], very small mobilities were observed. In studies starting from very low concentrations, the mobility was found to first increase, but then to saturate with increasing particle number density, $n$ (related to the volume fraction, $\Phi$, as:

$n = \Phi / V_P = 3\Phi / 4 \pi a^3$ where $a$ is the particle radius) [24, 25, 26]. There, the transition to a concentration independent mobility seems to correlate to the onset of inter-particle ordering, i.e. to conditions of strongly overlapping electric double layers. Moreover, increase and saturation have been observed in systems of quite different salt concentrations $10^{-6}$ mol l$^{-1}$ ≤ $c_S$ ≤ $10^{-2}$ mol l$^{-1}$ [27]. The observed behavior stays theoretically unexplained. However, in other studies we observed the mobility to start with a plateau, but then to decrease [28, 29]. In the latter case, the onset of the decrease coincides with the onset of counter ion dominance, i.e. when the counter-ions of the charged particles start to dominate the small ion population [30]. This is reproduced in computer simulations [29] and theoretically understood in terms of the electrolyte concentration dependence of the zeta potential [31]. Together, the experimental observations seem to suggest the existence of a broad maximum in the density dependent mobility [30]. However, so far, no single study has reported all the three different dependencies for a given particle species, which would resolve the apparent contrast between different observations.

The reason for this lack of data is mainly technical. Practically all data showing the ascent were measured on particles in aqueous suspension with large scattering cross section, i.e. particles with large radii, $a$, and/or refractive index contrast. This yields a good signal-to-noise-ratio even at low particle number density, $n$, but leads to problems with MS contributions at large $n$, where the descent occurs. Conversely, small particles are causing less MS and can be investigated up to large $n$ and even in the crystalline region of the phase diagram [32]. However, at low $n$ the signal becomes vanishingly small and the ascent is not accessible.

In the present paper, we therefore suggest a facile empirical correction applicable to moderate multiple scattering, which considerably extends the accessible range of particle number densities in the presently used LDV-setup. It is based on the use of a low angle scattering configuration, which restricts the detection volume, as well as the direction and magnitude of detected MS-scattering vectors. In our geometry, the remaining MS signal is well approximated in frequency space by a single Lorentzian. We exploit the observed pronounced difference in the relaxation time scales for multiply and singly scattered light. This allows subtracting the MS contribution to the Doppler spectra after a simple fit procedure and thus isolating the singly scattered spectral contribution. The isolated MS contribution shows a number of interesting features. It is a well-defined Lorentzian of finite width, which increases linearly with particle concentration over the range of $n$ studied. These observations may in future studies help to develop a heterodyne MS theory for light scattering in transmission geometry, which is presently lacking.

We demonstrate the scope and limits of our correction scheme by three examples. First, we utilize our approach in measurements of Doppler spectra in low salt suspensions of low to moderate turbidity under a Poiseuille-type electro-kinetic flow. The sample investigated consists of weakly ordering optically homogeneous but polydisperse latex spheres yielding a dominance of incoherent scattering and thus an independence of the signal from sample structure [19]. Using our empirical MS-correction scheme, we evaluate the isolated single scattering spectra for electro-kinetic velocities. We observe a broad maximum for $\mu_e$ at moderate particle number densities. Albeit still preliminary, our study thus indicates the possibility for a reconciliation of previously divergent interpretations of the concentration dependence of the electro-phoretic mobility in future systematic studies.

For this sample, we also obtain the effective diffusion coefficient, $D_{eff}$, with and without the presence of shearing flow. In the absence of shear, $D_{eff}$ stays close to the standard Stokes-Einstein-Sutherland diffusion coefficient, $D_0$. This is in accordance with the expectation for low salt charged sphere suspensions with dominant incoherent scattering and the very small volume fraction dependence of the there-measured self-diffusion coefficient, $D_S$ [33]. In the electro-kinetic experiments, $D_{eff}$ increases linearly with increasing field strength and increasing concentration of particles, but $D_0$ is recovered in the limit of zero shear and infinite dilution.

In a second set of experiments, we study suspensions of optically inhomogeneous co-polymer colloids of low polydispersity under conditions of strong mutual interaction. We demonstrate the possibility to follow the decrease of $D_S$ during solidification and annealing of grain boundaries in two crystallizing samples with different *n*.

Our measurements demonstrate the suitability of the empirical correction scheme for measurements of electro-kinetic and self-diffusive properties of multiply scattering suspensions without much extra effort as compared to standard low angle integral LDV. We anticipate that this approach may also turn out useful for other configurations used in dynamic light scattering.

In what follows we first outline the experimental procedure, sketching the integral low angle super-heterodyne LDV setup and the corresponding single scattering theory, introducing the samples, showing typical spectra and explain the novel correction scheme. The results section reports on the characterization of the MS signal contribution and presents MS corrected results of electro-kinetic and diffusion experiments. In the discussion section, we address the limits and scope of our approach, anticipate future applications and discuss the results obtained from MS-corrected data. We end with some short conclusions.

**EXPERIMENTAL METHODS AND EVALUATION SCHEME**

**Integral, low-angle, super-heterodyne LDV**

Our home-built Laser Doppler Velocimetry (LDV) instrument combines the reference beam integral measurement of [34] (allowing to measure the complete flow-profile in a single measurement) with super-heterodyning [35, 36, 37] (allowing to separate the heterodyne part of the power spectrum from the homodyne part, electronic and low frequency noise) and a significant restriction of the detection volume [15, 16] (to reduce detection of MS photons by approaching confocal detection conditions). The present machine is a modification of a previously described setup [19] now also allowing a continuous variation of the scattering angle. For additional technical details see supplementary materials. An essential part of this instrument is the detection geometry sketched in Fig. 1. Additional details are given in the supplementary materials (Fig. S1 and S3).

The (single) scattering vector $\mathbf{q}_1 = \mathbf{k}_i - \mathbf{k}_1$ (where $\mathbf{k}_i$ and $\mathbf{k}_1$ are the wave vectors of the illuminating and one-time scattered beams) is parallel to the applied field direction (c.f., Fig. S1). In the present sign convention, $\mathbf{q}$ is proportional to the momentum transfer from the photon to the scattering particle. Its modulus is given by $q = (4\pi \upsilon_S / \lambda_0) \sin(\Theta_S/2) = (4\pi / \lambda)$ $\sin(\Theta/2)$, where $\upsilon_S$ is the index of refraction of the solvent and $\Theta_S = 6°$ is the beam crossing angle in suspension. The electric field points downward in $-z$ direction (c. f. Fig. S1). We note that light scattered by a particle $j$ moving with a velocity $\mathbf{v}_j$ in z-direction is Doppler-shifted by the circular frequency $\omega_D = -\mathbf{q}\cdot\mathbf{v}(x,y)$, where the velocity can be a function of the position. The frequency shift, $\omega_D$, is positive, if the particle velocity shares an obtuse angle with the scattering vector, $\mathbf{q} = q\,\hat{\mathbf{z}}$, where $\hat{\mathbf{z}}$ is the unit vector in z-direction. Hence, a positive Doppler shift corresponds to particles moving towards the detector with a negative velocity with respect to the field direction.

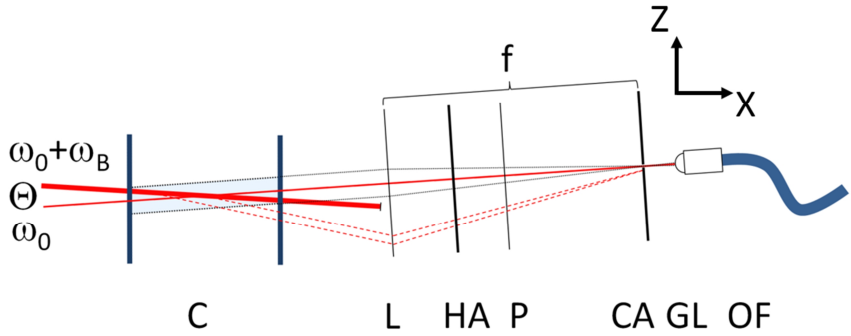

**Fig. 1:** (color online) Sketch of the detection geometry. Shown is a cut in the scattering plane (x-z). In the cell (C) containing the suspension, the frequency shifted illumination beam (thick red line) is crossed by the reference beam (thin red line) under an angle $\Theta_S$. (Note that for simplicity we here only show light paths in air with $\Theta = 8°$). The illumination beam is stopped by a beam stopper. Light leaving the cell parallel to the reference beam is focused by the lens (L, f = 50mm) into the focal plane, where a circular aperture (CA) is placed. The position and width of CA defines the direction of the scattering vector $\mathbf{q}$ and the q-resolution $\Delta q$, respectively. The minimum opening is chosen to about 1mm in order let a sufficient amount of reference beam light pass. A grin lens (GL, diameter 4mm) mounted on an optical fiber (OF) is aligned collinear with the reference beam and guides the collected light to the detector. The distance between CA and GL defines the diameter of the detection volume (light grey). It is chosen as to contain the complete illuminating beam crossing the cell. Light originating outside the detection volume as well as light leaving the cell in directions not sufficiently parallel to the reference beam (e.g. dashed red lines) is blocked. Light scattered out of the x-z plane is optionally rejected by a horizontally mounted slit aperture (HA). A polarizer (P) assures V/V detection. If L and CA are removed, the q-resolution is defined by solely the acceptance angle of the grin lens and the maximum detection volume by the acceptance angle and the GL diameter. It typically is larger than with CA and L and is slightly conical.

The reference beam is also directed into the fiber and is superimposed with the scattered light. It therefore acts as a local oscillator, and gives rise to beats in the intensity observed at the detector. These are analyzed by a Fast Fourier Transform analyzer (Ono Sokki DS2000, Compumess, Germany) to yield the power spectrum as a function of frequency, $f = \omega / 2\pi$. To obtain a good signal to noise ratio typically some 200-2000 subsequent spectra are averaged. The power spectra are transferred to a computer, where the MS correction is applied and the corrected spectra are further evaluated for the electro-kinetic mobilities.

**Single scattering super-heterodyne Doppler spectra**

Some progress has been made in the treatment of double scattering in homodyne experiments [13, 38, 39]. Still, the complexity of theory for triple and higher-order scattering means, that the theoretical approach has found little application in practice. Multiple scattering has been treated for heterodyne low coherence backscattering experiments [17, 18], but a general theory of MS scattering for (super-)heterodyne techniques is not known to us. We therefore correct our experimental signals for noise background and MS contributions (see below) and apply single scattering theory for their evaluation. (Super-)heterodyning is based on the detection of Doppler shifts imposed on the scattered photons at the instant of scattering and therefore does not suffer from de-coherence of scattered beams due to changes of the relative distance between scatterers. Our theoretical frame for single scattering is based on earlier work on homo- and heterodyning techniques in dynamic light scattering [11, 40]. A theory of conventional heterodyne LDV using an integral reference beam set-up was outlined in [40]. Super heterodyne theory for integral measurements at low angles has been detailed in [19]. Further extensions to include de-correlation effects such as Taylor-dispersion arising in sheared solvents [41] and a rigorous treatment of electro-phoretic mobility polydispersity will be presented in a forthcoming paper [42]. Here, we only summarize the main theoretical expressions for single scattering following [19]. Some additional details are given in the supplementary material.

The power or Doppler spectrum $C_{shet}(\mathbf{q},\omega)$, is the time Fourier transformation of the mixed-field intensity autocorrelation function, $C_{shet}(\mathbf{q},\tau)$:

$$C_{shet}(\mathbf{q},\omega) = \frac{1}{\pi}\int_{-\infty}^{\infty} d\tau \; \cos(\omega\tau) C_{shet}(\mathbf{q},\tau) \tag{1}$$

with circular frequency ω and correlation time τ. For a scattered light field of Gaussian statistics the latter quantity is given as:

$$C_{shet}(\mathbf{q},\tau) = \left(I_r + \langle I_1(\mathbf{q})\rangle\right)^2 + 2I_r\langle I_1(\mathbf{q})\rangle \mathrm{Re}\left[\hat{g}_E(\mathbf{q},\tau)\exp(-i\omega_B\tau)\right] + \langle I_1(\mathbf{q})\rangle^2 |\hat{g}_E(\mathbf{q},\tau)|^2 \tag{2}$$

Here, $I_r$ is the reference beam intensity, and $\langle I_1(\mathbf{q})\rangle$ is the time-averaged singly scattered intensity, and $\hat{g}_E(\mathbf{q},\tau) = g_E(\mathbf{q},\tau)/\langle I_1(\mathbf{q})\rangle$ is the normalized field autocorrelation function. For homogeneous suspensions of interaction monodisperse, optically homogeneous, mono-sized particles, $\langle I_1(\mathbf{q})\rangle$ is given by:

$$\langle I_1(\mathbf{q})\rangle = I_0 n b^2(0) P(q) S(q) \tag{3}$$

Here, $I_0$ is a constant comprising experimental boundary conditions like illuminating intensity, distance from the sample to the detector, polarization details. $b^2(0)$ is the single particle forward scattering cross section, n is the particle number density, P(q) denotes the particle form factor and S(q) the static structure factor [39].

We assume the particles to undergo Brownian motion with an effective diffusion coefficient $D_{eff}$ and directed motion with a constant and homogenous drift with velocity $\mathbf{v}_0$. For this simple case indicated by the index 0, the power spectrum reads

$$\begin{aligned}
C_{shet}(\mathbf{q},\omega) = & \left[I_r + \langle I_1(\mathbf{q})\rangle\right]^2 \delta(\omega) \\
& + \frac{I_r \langle I_1(\mathbf{q})\rangle}{\pi}\left[\frac{q^2 D_{eff}(q)}{(\omega+[\omega_B-\omega_D])^2 + (q^2 D_{eff}(q))^2} + \frac{q^2 D_{eff}(q)}{(\omega-[\omega_B-\omega_D])^2 + (q^2 D_{eff}(q))^2}\right] \\
& + \frac{\langle I_1^0(\mathbf{q})\rangle^2}{\pi}\frac{2q^2 D_{eff}(q)}{\omega^2 + (2q^2 D_{eff}(q))^2}
\end{aligned} \tag{4}$$

$D_{eff}$ depends on q, in case collective diffusion is measured, and $\omega_D = \mathbf{q}\cdot\mathbf{v}_0$ is the frequency of the resulting Doppler shift. The spectrum contains three contributions: a trivial constant term centered at zero frequency, two super-heterodyne Lorentzians of spectral width $q^2 D_{eff}$ shifted away from the origin by both the Bragg frequency and the Doppler frequency, and the homodyne Lorentzian of double width which is independent of the particle drift motion and is again centered at the origin. The homodyne term is known to be seriously affected by shear [43] and multiple scattering resulting in the loss of coherence of the scattered light, which renders it ill-defined. From Eqn. (4), we note that the desired information about the electro-phoretic and diffusive particle motion is fully contained in each of the super-heterodyne Lorentzians which are symmetric about the origin. In Fig. 4, and also the following, we therefore display the measured data only for positive frequencies centered about the positive Bragg shift frequency. Interestingly, this is already sufficient to infer the sign of the electro-kinetic velocities. (The complete positive part of a complete Doppler spectrum is displayed in Fig. S2 of the supplementary material). According to our sign convention, a positive center of mass frequency corresponds to a negative electro-phoretic velocity $\mathbf{v}_{ep}$ and hence a negative electro-phoretic mobility $\mu_{ep}$ and a negative particle charge. A maximum at the high frequency side corresponds to a positive wall velocity $\mathbf{u}_{eo}$ and hence to a negatively charged wall.

Under inhomogeneous flow, particles show a normalized velocity distribution, $p(v)$. The Doppler spectra then display a corresponding normalized distribution of Doppler frequencies, $p(\omega_D)$. This can be accounted for by a convolution integral:

$$C_{shet}(\mathbf{q},\omega) = \int d\omega_D \; p(\omega_D) C^0_{shet}(\mathbf{q},\omega) \qquad (5)$$

Most notably, structure may couple to flow properties resulting in inhomogeneous shear thinning or shear banding [44, 45], alteration of the particle flow profile due to local crystal cohesion [46, 47], and even time-dependent effects [48]. In such cases, where $p(\omega_D)$ is not known, only the electro-phoretic velocity can be obtained as the $y$-average of the $x$-averaged particle velocities $v_{ep} = \langle\langle v_P \rangle_x \rangle_y$ [48]. A suitable expression of $p(v)$ for Poiseuille-type electro-osmotic electro-phoretic flow is given in Eqn. (S1) and displayed in Fig. S3 in the supplementary material.

In previous studies, it was further observed that flow may couple to the diffusivity, and a dependence of fitted $D_{eff}$ on the field strength has frequently been observed and in most cases was found to be linear [34]. Underlying reasons could be Taylor dispersion [41] accounting for effects of shear, field induced velocity fluctuations [12] or a polydispersity of the electro-phoretic mobility [49]. In the present study, we fit measured spectra using an effective diffusion coefficient $D_{eff}(E)$, but only in the field free case or for the values obtained by extrapolation to $E = 0$, an interpretation in terms of a diffusion coefficient is attempted.

For truly mono-sized, optically homogeneous particles, only collective diffusion is measured, with $D_{eff} = D_C(q)$ depending on the scattering vector through hydrodynamic and direct particle interactions. The presently used samples in addition show a significant amount of depolarized scattering due to MS and incoherent scattering from individual particles differing in size or being optically inhomogeneous. The presence of these contributions can be noted already from visual inspection (c. f. Fig. S4 in the supplementary material) and influence on the power spectrum. For optically homogeneous particles with a finite size polydispersity of standard deviation $s$, the field autocorrelation function can be written using the decoupling approximation as [19]:

$$g_E(\mathbf{q},\tau) = I_0 n \overline{f^2(q)}\, S_M(\mathbf{q},\tau) \approx I_0 n \overline{f^2(q)} \left[ 9s^2 G(q,\tau) + S(\mathbf{q},\tau) \right] \qquad (6)$$

The corresponding normalized field autocorrelation function $\hat{g}_E(\mathbf{q},\tau) = g_E(\mathbf{q},\tau) / I_0 n \overline{f^2(q)}$ $S_M(\mathbf{q})$ is then used in Eq. (2). Here, $\overline{f^2(q)}$ is the average scattering cross section of the particles. $S_M(\mathbf{q},\tau)$ is the so-called measurable dynamic structure factor, $S(\mathbf{q},\tau)$ is the dynamic structure factor or intermediate scattering function arising from coherent scattering and $G(q,\tau)$ is the self-intermediate scattering function. The amplitude of the latter for $\tau = 0$ is independent of the suspension structure and the chosen $\mathbf{q}$ [51]. Note, that Eqn. (6) describes a limiting case of size-polydispersity. Any additional optical polydispersity, e. g. due to internal inhomogeneity of the particles further increases the weight of $G(q,\tau)$.

With evolving structure, coherently scattered and incoherently scattered contributions to the spectra will differ. The difference is most pronounced at small $q$, where $S(q)$ becomes very small due to the decreased isothermal compressibility of the suspension [50, 51]. For charged particles with large optical polydispersity under low salt conditions this may even lead to a dominance of the incoherent scattering contribution at small angles which was previously utilized to measure electro kinetic properties of ordered colloidal fluids and crystals showing negligible MS [32]. It is here used to study self-diffusion using $D_{eff} = D_S$. The field free self-diffusion coefficient is expected to scale with the volume fraction $\Phi$ as $D_S = D_0 (1-a_s\Phi^{4/3})$ with $a_s$ decreasing with increasing and interaction strength [33]. The opposite limit of a vanishing incoherent contribution in which collective properties are measurable would be obtained for suspensions of optically homogeneous particles of low polydispersity, but it was not investigated here.

**Samples and optical characterization**

Sample PnBAPS118 was employed for measurements of self-diffusion in crystallizing and coarsening colloidal solids. These particles are a 35:65 W/W copolymer of Poly-n-Butylacrylamide (PnBA) and Polystyrene (PS), kindly provided by BASF, Ludwigshafen (Lab code PnBAPS118 manufacturer Batch No. 1234/2762/6379). Their nominal diameter of (117.6 ±0.65) nm and nominal polydispersity index of PI = 0.011 were determined by the manufacturer utilizing dynamic light scattering. TEM analysis and form factor measurements using SAXS yield PI ≈ 0.06. The effective charge for PnBAPS118 is $Z_{eff}$ = 647±18. Due to the combined effect of optical inhomogeneity and moderate polydispersity, these particles show strong incoherent scattering.

Sample PS310 was used for measurements of electro-kinetic properties and self-diffusion with and without shearing flow. It comprises of commercial polystyrene latexes stabilized by carboxylate surface groups (lab code PS310, manufacturer batch #1421. IDC, Portland, USA). Their nominal diameter is 310 nm as given by the manufacturer. Their hydrodynamic radius $a_H$ = 167 nm and polydispersity index PI = s/<a> = 0.08 were determined by dynamic light scattering measurements on dilute samples [52]. Their effective charge is $Z_{eff}$ = 5260±100. Also these particles show a considerable amount of incoherent scattering due to size polydispersity. Typical distributions of singly scattered light in dependence on scattering angle are shown for the two investigated suspensions in Fig. S5a and b.

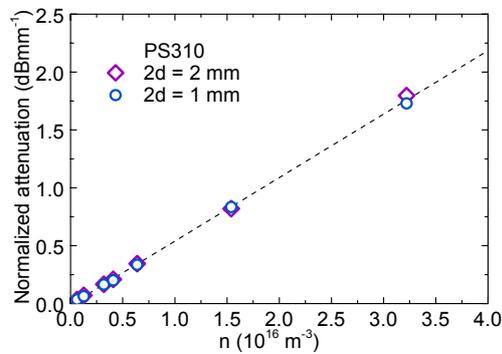

**Fig. 2:** (color online) Attenuation of transmitted intensity normalized to the cell thickness 2d as a function of the number density n. Results of the two measurements coincide. A fit of Eqn. (7) returns an attenuation cross-section of $\sigma_{633}$= 0.013±0.1 μm$^2$.

Suspensions of different *n* were first conditioned as described in the supplementary materials. Small amounts of the PS310 suspension were then pipetted to rectangular quartz cells of 2d = 1 mm and 2d = 2 mm optical path length (Hellma, Germany) and transferred to a home built turbidity measurement working at 633 nm. The transmittance or relative transmitted intensity, $T = I/I_0$, was recorded as a function of $n$, where $I_0$ refers to the cells filled with filtered milli-Q grade water. Fig. 2 shows the result. We note that conductivity measurements are not feasible in this set-up. Therefore, samples were exposed to ambient air to obtain $CO_2$-saturation. Here, the residual ion concentration was estimated to be $c_S \leq 5 \times 10^6$ mol l$^{-1}$.

The transmitted intensity for deionized or low salt suspensions is known to obey the Lambert-Beer-law in both the fluid and crystalline state [53]:

$$2dn\sigma_{633} = -\ln\left(\frac{I}{I_0}\right) \tag{7}$$

From fits of Eqn. (7) to the data, we obtained an attenuation cross-section of $\sigma_{633} = 0.013 \pm 0.1$ μm$^2$.

It is paramount to ensure identical optical properties in both turbidity experiments and LDV. PS310 were therefore conditioned in the same way in both experiments. The background salinity in addition is advantageous for the goals of the present LDV study. At the volume fractions utilized, the suspensions will develop only a weak fluid order and a pronounced forward scattering will still be present. This is ideally suited to demonstrate the scope of the here introduced MS-correction scheme.

For each concentration, a second amount of sample was therefore filled into the electro-phoresis cell (Standard EL10 by Rank Bros., Bottisham, Cambridge, UK or replica by Lightpath Optical Ltd., Milton Keynes, UK) used in the SH-LDV experiment. The u-shaped electro-phoretic cell has its electrode chambers separated from the actual optical section of rectangular cross section of (10×1) or (10×2) mm$^2$ and width $l$ = 40 mm. The effective platinum electrode distance is $L \approx 80$ mm. It was determined precisely before each measurement series from calibration with an electrolyte dilution series. During the electro-kinetic measurements, the electrode chambers are sealed. An alternating square-wave field of strength up to $E_{MAX} = U/L = 15$ V cm$^{-1}$ was applied. To avoid accumulation of particles at the electrodes and to further ensure fully developed stationary flows [35], field switching frequencies $f_{AC} = (0.02 - 0.1)$ Hz were used. Measurement intervals were restricted to one field direction. Each was starting after the full development of electro-osmotic flow profile and ending shortly before the field reversal. For each field strength, we recorded and averaged 250-400 full time-frame power spectra. For diffusion measurements in PnBAPS118, no field was applied. We here typically averaged over 1000-2000 time frames. For rapidly crystallizing samples of that species, many repeated shear-melting - recrystallization cycles were needed to obtain the temporal evolution of the diffusive properties.

**Typical spectra and fitting procedure**

Typical examples of averaged power spectra for PS310 and $E = 0$ at $n = 1.64\times10^{16}$ m$^{-3}$ as well as for $E = 4.3$ V cm$^{-1}$ and increasing n are shown in Fig. 3 and 4, respectively. Fig. 3 shows that without applied field, only a single broad spectral feature centered at 2 kHz is present. Raw spectra for PnBAPS118 appear to be similar at smaller overall spectral power. The singular feature varies in strength and broadness for changing $n$.

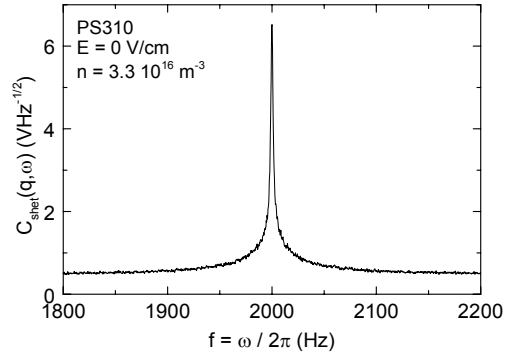

**Fig. 3:** Power spectrum of PS310 at $n = 1.64\times10^{16}$ m$^{-3}$ and $2d = 2$ mm, recorded in the multiple scattering regime without applied electric field. A single broad spectral feature is visible which is centered at the Bragg shift frequency, $f_B = 2000$ Hz.

This is different in Fig. 4 for the case of an applied field, where we can clearly discriminate different types of spectral contributions due to their different spectral shape. At very low n, the characteristic shape of the electro-phoretic signal [21] is barely visible in the background noise. With extensive further averaging, an acceptable signal may be obtained for n ≥ $6\times10^{14}$ m$^{-3}$. Excellent low noise signals are obtained for $9\times10^{14}$ m$^{-3}$ ≤ $n$ ≤ $8\times10^{15}$ m$^{-3}$. A broad additional signal component becomes clearly discernible for $n ≥ 1\times10^{16}$ m$^{-3}$. It increases with increasing $n$ until it dominates the power spectrum for $n > 10^{17}$ m$^{-3}$. It appears to be symmetric with a maximum close to the Bragg frequency, $f_{MS} \approx f_B = 2000$Hz. As will be shown below, it can be demonstrated to originate from multiple scattering. At $n = 1\times10^{17}$ m$^{-3}$, the sample transmission T < 0.1, and also the reference beam is strongly attenuated. The spectrum becomes noisier and the signal to background ratio decreases significantly. Finally, all super-heterodyne signal contributions disappear, and the spectrum becomes finally dominated by electric power grid harmonics.

For evaluation of spectra taken at low densities which are free of the extra signal contribution, we average the noise background at frequencies far off the other spectral features and subtract it from the data. We then perform a least square fit of the SS theoretical expression for $C_{shet}(q,\omega)$ derived from a combination of Eqns. (1) to (5) and Eqn. (S1) with the cell geometry, the field strength and the particle number density as an input. For spectra in the range of $8\times10^{14}$ m$^{-3}$ ≤ $n$ ≤ $5\times10^{15}$ m$^{-3}$, we obtained excellent fits with $u_{eo}$, v and $D_{eff}$ as independent fit parameters. However, attempts to fit the spectrum at larger $n$ lead to unsatisfactory results due to the contribution by multiple scattering (MS). This is illustrated in Fig. 6a for a spectrum recorded on PS310 at deionized conditions, $E$ = 5.3 V cm$^{-1}$, $n$ = $1.64\times10^{16}$ m$^{-3}$ and $2d$ = 2 mm.

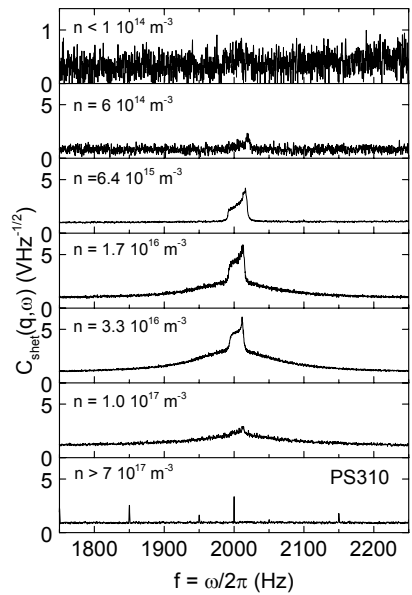

**Fig. 4**: Examples of power spectra obtained at a field strength of E = 4.3 V cm$^{-1}$ for $CO_2$-saturated suspensions of PS310 at different particle number densities as indicated. Background noise, electro-phoretic spectrum and the broad MS peak each show distinct spectral shapes. The ratio of these contributions changes with increasing n. For PS310, the actual Doppler-spectra emerge from the background noise for $n > 5\times10^{14}$ m$^{-3}$ and vanish in the broad feature for $n > 1\times10^{17}$ m$^{-3}$. Both contributions are centered close to the Bragg shift frequency, $f_B$ = 2000 Hz. In the limiting case of optically dense samples, transparency is lost

and the signal disappears completely leaving only the contributions of electronic noise. Note that the characteristic Doppler spectrum is clearly visible for concentrations covering more than two orders of magnitude in n.

**Empirical correction procedure**

We devised a simple empirical scheme that can be applied for the correction of MS contaminated data. We first address the field free case (c.f. Fig. 5). We assume a superposition of different, statistically independent spectral contributions as described by:

$$C_{shet}(\mathbf{q},\omega) = A_1 + A_2 \int d\omega \ p(\omega) C^0_{shet}(\mathbf{q},\omega,v,D_{eff}) + \frac{A_3}{2\pi} \frac{w_{MS}}{(\omega-\omega_0)^2 + w_{MS}^2} \qquad (8)$$

Here, $A_1$ is a constant describing the frequency independent noise background, $A_2$ is the integrated spectral power of the Doppler spectrum depending on $I_r<I_1(\mathbf{q})>$, and $A_3$ is integrated spectral power of the MS contribution depending on $I_r<I_N(\mathbf{q})>$, where N denotes the number of scattering events. The MS term is modelled as Lorentzian centered at $\omega_0$ and of full width at half height, $w_{MS}$ [18, 54]. For $E = 0$, $v = 0$ and only Eqn. (1) to (4) are necessary to calculate the single scattering super-heterodyne spectrum. An example fit to data obtained in the field free case is shown in Fig. 5, with the inset magnifying the central region of the spectrum. An excellent fit is obtained, showing the applicability of our empirical correction scheme.

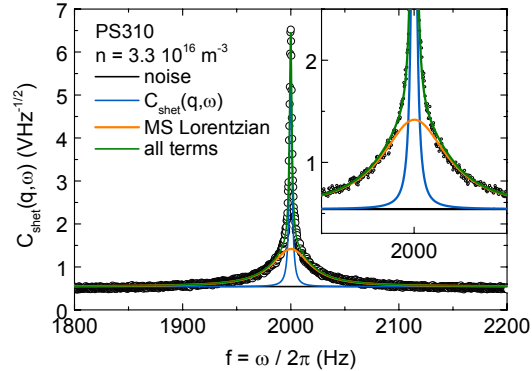

**Fig. 5**: (color online) Power spectrum of PS310 at $n = 1.64 \times 10^{16}$ m$^{-3}$ and $2d = 2$ mm, recorded in the multiple scattering regime at T = 0.4 without applied electric field (black circles). Also shown is the fit of Eqn. (8) (solid green line) which is the superposition of a constant background noise (solid black line), the Lorentzian single scattering super-heterodyne power

spectrum (solid blue line) and a second Lorentzian describing the multiple scattering contribution (solid orange line). Inset: magnification of the central region of the signal.

Next, we applied the correction scheme to data obtained at finite field strengths. In Fig. 6b we show the background noise corrected spectrum with a Lorentzian fitted (orange solid curve) to the wings of the remaining MS spectral feature. After subtraction of the MS contribution, the remaining experimental data are fitted in Fig. 6c by a theoretical expression based on Eqn. (1) to (5) and (S1) (blue curve). An excellent fit is obtained for the fit parameters:

$v_{ep}$ = 37.4 μm s$^{-1}$, $u_{eo}$ = 64.1 μm s$^{-1}$, $D_{eff}$ = 4.49×10$^{-12}$ m$^2$ s$^{-1}$.

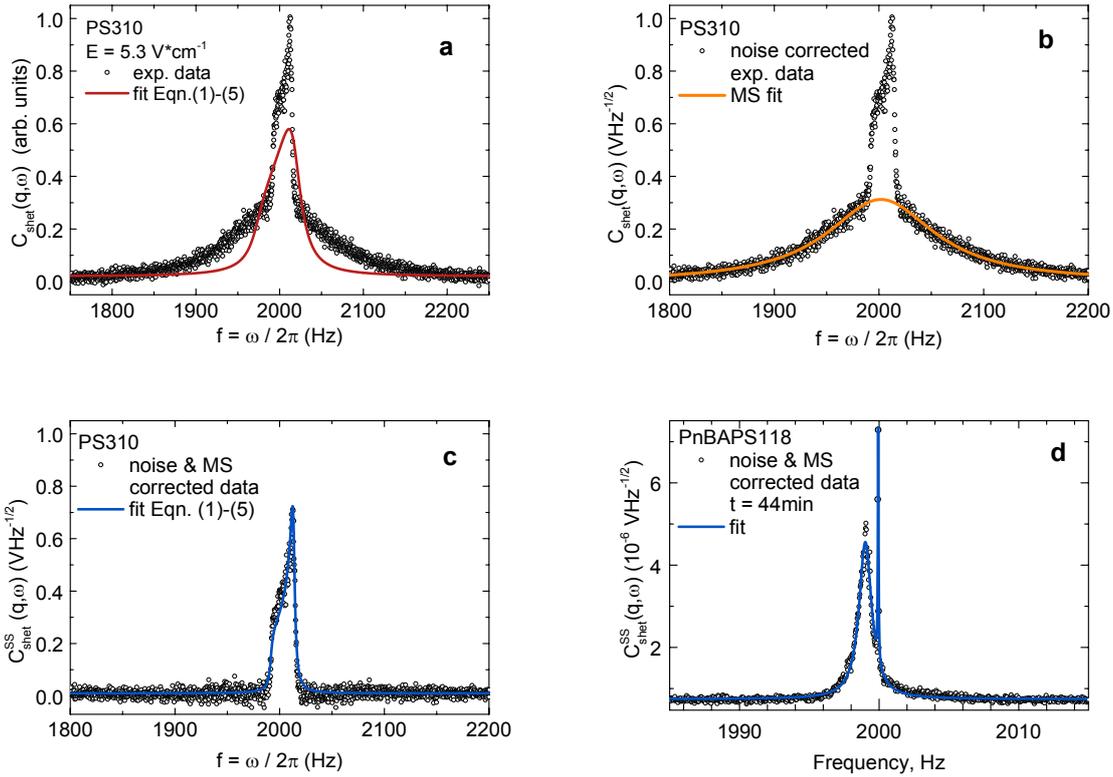

**Fig. 6**: (color online) Fitting of data recorded with applied field. a) Uncorrected low angle integral super-heterodyne spectrum of PS310 recorded at $E$ = 5.3 V cm$^{-1}$ with $n$ = 1.64×10$^{16}$ m$^{-3}$, cell depth $2d$ = 2 mm and cell height $2h$ = 10 mm. Data are normalized to unity at the maximum. The red solid line shows the best least square fit obtainable using Eqn. (1)-(5) and (S1) b) Experimental data of a) after subtraction of the small constant noise background term. The MS Lorentzian (orange solid line) is fitted to the wings of the spectral feature. c) Isolated single scattering power spectrum obtained by subtraction the fitted MS contribution from the data in b). The single scattering data are fitted with the theoretical expression for the single scattering integral super-heterodyne power spectrum based on Eqns.

(1) to (5) and (S1). The fit returns $v_{ep}$ = 37.4 µm s$^{-1}$, $u_{eo}$ = 64.1 µm s$^{-1}$, $D_{eff}$ = 4.49×10$^{-12}$ m$^2$ s$^{-1}$. d) Isolated single scattering power spectrum of poly-crystalline PnBAPS118 at $n$ = 4.47×10$^{18}$ m$^{-3}$ obtained after noise and MS correction. Note the low signal intensity as compared to Fig. 6. This spectrum was recorded 44 min after the stop of the shear over a time interval of 5 min. A minute plug-flow was applied to separate the spectral contributions (for details see text). Data are fitted with a theoretical expression for the single scattering integral super-heterodyne power spectrum based on Eqns. (1) to (5) plus an additional Lorentzian for the 2 kHz peak. The fit returns $D_S$ = 1.43×10$^{-13}$ m$^2$ s$^{-1}$.

For PnBAPS118, MS contribution, background noise and SS signal appeared to be much weaker than for PS310 even at large number densities. Still, MS correction was both necessary and unproblematic. In some cases, however, the single scattering spectrum was so weak that it became comparable to a residual 2 kHz feature. The latter originates from the interference of the reference with of parasitic reflections from the illumination beam and is not considered in Eqn. (4). We therefore applied a minute plug flow (v ≈ 1 µm s$^{-1}$) to shift the centre of the heterodyne Lorentz by a few Hz and separate the signal from the artefact. This is shown in Fig 6d. We then fitted the spectrum by our expression for the single scattering integral super-heterodyne power spectrum based on Eqns. (1) to (5) plus an additional narrow Lorentzian for the 2 kHz-peak (blue curve). An excellent fit is obtained returning $D_S$ = 1.43×10$^{-13}$ m$^2$ s$^{-1}$.

**RESULTS**

**The multiple scattering contribution**

We performed several tests of the origin and behavior of the different signal contributions. These are shown in the supplementary materials (Figs. S7 to S9) and discussed in detail below. From these, we attribute the broad-peaked spectral contribution appearing at elevated particle concentrations to multiple scattering. The super-heterodyne spectra taken in the frequency range close to $f_B$ are thus composed of a background resulting from noise, a Doppler spectrum and a symmetric MS background (Fig. S7a). Integrated spectral powers for noise background, SS and MS spectral contributions all scale with the reference beam intensity (c. f. Fig. S7b).

The MS contribution can be excellently fitted by a single Lorentzian and characterized by its width, $w_{MS}$, integrated spectral power, $A_3$, and the position of its center of mass, $\omega_0$. The statistical errors are small for $w_{MS}$ and $A_3$, but somewhat larger for $\omega_0$ due to the restriction of the fit range to spectral regions outside the SS-signal. $A_3$ is reliably characterized for $n \geq 10^{16}$ $m^{-3}$. $A_3$, $A_2$ and $A_1$ are related linearly to the reference beam intensity (Fig. S7b). Further, in plots of the dependence of the integrated spectral powers versus number density, the scatter in the absolute values of $A_3$ and $A_2$ is highly correlated to the scatter in $A_1$ (see Figs. S8 and S9a and b). This highly non-reproducible but systematic scatter is attributed to inevitable slight differences in cell alignment after each change of sample leading to slightly different reference beam intensities at the detector. Therefore, we show these data in terms of the noise normalized integrated spectral power $A_i^* = A_i / A_1$.

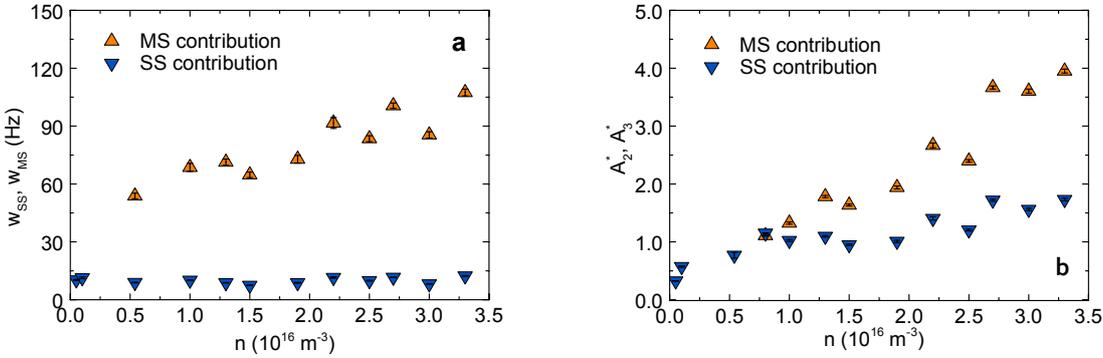

**Fig. 7**: (color online) MS and SS characteristics obtained from fits of a Lorentz function to the MS (up triangles) and fits of Eq. (1) to (5) to the SS (down triangles) contributions as measured in field free samples of PS310 at different number densities $n$. a) Spectral width, $w_{MS}$ and $w_{SS} = D_{eff}q^2$. b) Noise normalized integrated spectral power. In order to minimize effects of altered reference beam detection conditions after sample exchange, we show the data in terms of the noise normalized integrated spectral power $A_i^* = A_i / A_1$.

Fig. 7a shows that $w_{MS}$ increases linearly starting from a finite value. By contrast, $w_{SS} = D_{eff}q^2$ used in fits of Eq. (1) to (5) for the SS contribution stays constant at a low value. Fig. 7b compares the zero field integrated spectral power of the SS contribution to that of the MS contribution. The single scattering $A_2^*$ increases monotonously with decreasing slope. The multiple scattering $A_3^*$ increases with a constant slope.

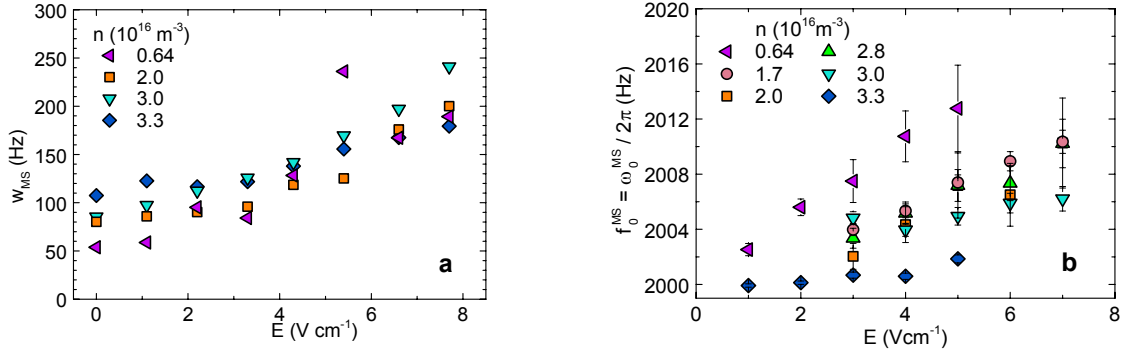

**Fig. 8:** (color online) a) Width of the MS contribution as a function of applied field strength for different number densities as indicated. We observe a roughly linear increase from an n-dependent offset at $E = 0$ V cm$^{-1}$. b) Central frequencies of the MS Lorentzian as a function of applied field strength for different number densities as indicated.

Upon the application of an electric field, the MS-Lorentzian broadens approximately linearly with increasing field strength. Measured widths start from a small *n*-dependent offset of a few tens of Hz and increase approximately linearly up to values between 200 and 300 Hz at $E = 7.5$ V cm$^{-1}$. This is shown in Fig. 8a. The MS contribution is shifted towards larger frequencies with increasing field. Fig. 8b show data for the central frequency for six *n*-series. For each *n*, data assemble on approximately straight lines with lesser slopes for the larger *n*.

Spectra of PnBAPS118 were in general found to be much less affected by MS. However, when the turbidity was adjusted to large values by using ($10^3$-$10^4$)-fold larger number densities, the behavior of the MS contribution for PnBAPS118 was qualitatively similar to that of PS310. An example obtained at $n = 2\times10^{20}$ m$^{-3}$ is shown in Fig. S10 of the supplementary material. The MS contribution characteristics therefore appear to be independent of particle size and structure. The samples, of course, differ in their physical quantities due to different electro-kinetic properties, sample structure, diffusivity and interaction strength. This is revealed in the SS contributions and is discussed next. However, little of that information is retained in the MS signal.

**Evaluation of the isolated single scattering power spectrum**

<u>Electro-kinetics</u>

Main goal of this study was to find and apply a suitable MS-correction scheme in order to measure electro-kinetic velocities over an extended range of particle concentrations. In our measurements on deionized but approximately $CO_2$-saturated conditions, we measured five different field strengths per number density. For each parameter set, 250 spectra were averaged. An example of the obtained high quality MS free single scattering super-heterodyne power spectra is displayed in Fig. 9a. For this sample at $n = 3.3 \times 10^{16}$ m$^{-3}$, the transmittance was merely 0.4. As the field strength was increased, the spectra stretched and their centers of mass shifted to larger frequencies. This is excellently described by the corresponding fits, which were performed using a combination of Eqns. (1) to (5). They return the electro-phoretic velocity of the particles, $v_{ep}$, the electro-osmotic solvent velocity at the cell walls, $u_{eo}$, and an effective diffusion coefficient, $D_{eff}$. The obtained field-dependent velocities are displayed in Fig. 9b. They are on the order of several tens of μm s$^{-1}$ and show a strictly linear dependence on the applied field strength.

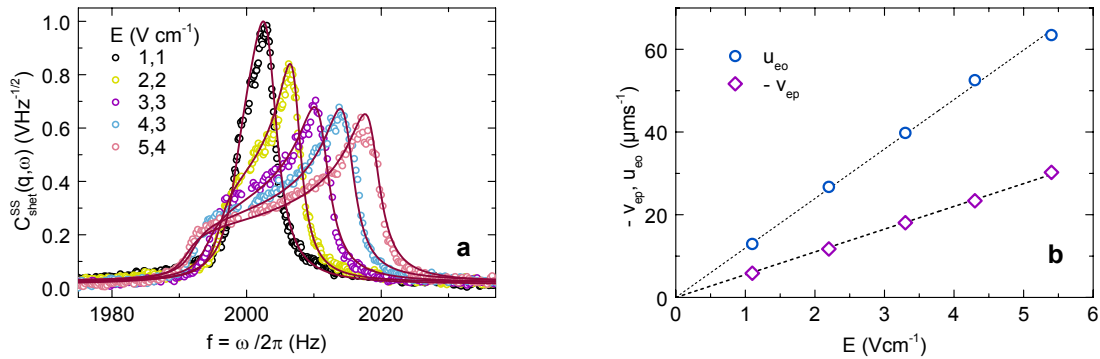

**Fig. 9**: (color online) a) Isolated single scattering Doppler spectra for PS310 at $n = 3.3 \times 10^{16}$ m$^{-3}$ and various field strengths as indicated. Solid lines are fits based on Eqn. (1) to (5). b) Field strength dependence of the electro-osmotic velocity at the cell walls, $u_{eo}$, and the negative of the electro-phoretic particle velocity, $-v_{ep}$, as obtained from the fits in a). The field dependence of both velocities is strictly linear. From the slope, we obtain the electro-phoretic and electro-osmotic mobilities: $\mu_{ep} = -5.4 \pm 0.1\ 10^{-8}$ m$^2$V$^{-1}$s$^{-1}$ and $\mu_{eo} = 1.2 \pm 0.2\ 10^{-7}$ m$^2$V$^{-1}$s$^{-1}$.

Under this condition of linear electro-kinetic response, the respective slopes equal the electro-phoretic particle mobility, $\mu_{ep}$, and the electro-osmotic mobility along the cell wall, $\mu_{eo}$. Following the IUPAC convention, both mobilities are converted to reduced units to eliminate the dependencies on solvent viscosity, η, and solvent dielectric permittivity $\varepsilon_0\varepsilon_r$ as well as on temperature, T [20]:

$$\mu^{red} = \mu \frac{3\eta e}{2\varepsilon_0 \varepsilon_r k_B T} \tag{9}$$

Here $e$ is the elementary charge and $k_B T$ denotes the thermal energy. The results for $\mu_{ep}^{red}$ are plotted in Fig. 10 as a function of particle number density. The reduced mobility first increases, then displays a plateau at a value of about 5.8, then decreases again. The corresponding reduced electro-osmotic mobilities are systematically larger, show a somewhat larger scatter due to wall coating effects and a much less pronounced density dependence.

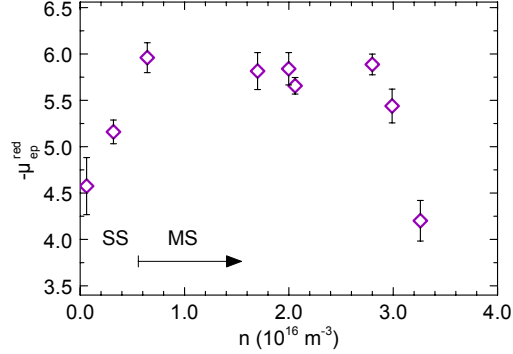

**Fig. 10:** (color online) Reduced electro-phoretic mobilities $\mu_{ep}^{red}$ for PS310, measured under deionized conditions as a function of the particle number density $n$ over roughly two orders of magnitude in $n$. Data for $n > 5 \times 10^{15}$ m$^{-3}$ were obtained performing both a correction for noise and for the MS scattering contribution. Note that the mobilities are negative for the negatively charged particle species. With increasing particle concentration, the electro-phoretic mobility shows an increase followed by a broad maximum and a steep decrease. The arrow marks the range for which MS-correction was necessary to obtain reliable fits of the electro-kinetic spectra.

Diffusion

While our experiment was not (yet) optimized for diffusion measurements, we nevertheless took the chance to check the performance of our correction scheme for such investigations. At low salt conditions PS310 shows weak ordering at elevated $n$. The SS-signal is dominated by incoherent scattering relating to the self-intermediate scattering function $G(q, \tau)$ and the effective self-diffusion coefficient, $D_S(q)$.

We fitted the noise and MS corrected spectra of PS310 by Eqn. (1) to (5) and S1. $D_{eff}(E)$ obtained at different concentrations are displayed in Fig. 11a a as a function of applied field strength. For each $n$, $D_{eff}$ depends linearly on E. Following [34], we performed least square linear fits to the data which return a slope that decreases with increasing $n$ and zero field extrapolated values of $D_{eff}(E \to 0)$ that increase with increasing $E$. This is shown in Fig. 11b. The offset increases approximately linear with increasing n starting from a value of $D_{eff}(n=0) = (1.23\pm0.4)\times10^{-12}$ m$^2$s$^{-1}$.

We further performed field free experiments and fitted the data using $v_{ep} = u_{eo} = 0$ m s$^{-1}$. Fig. 11b shows that within experimental uncertainty, $D_{eff} = D_S$ is independent of $n$ at a value of $D_S = (1.25\pm0.3)\times10^{-12}$ m$^2$s$^{-1}$. For comparison, the Stokes-Einstein-Sutherland self-diffusion coefficient for PS310 is $D_0 = k_B T/6\pi\eta a_H = (1.47\pm0.12)\times10^{-12}$ m$^2$s$^{-1}$, where we used a viscosity of water of 0.89 mPas, T=298.15 K and $a_H = 167$ nm as determined by standard Dynamic light scattering [52].

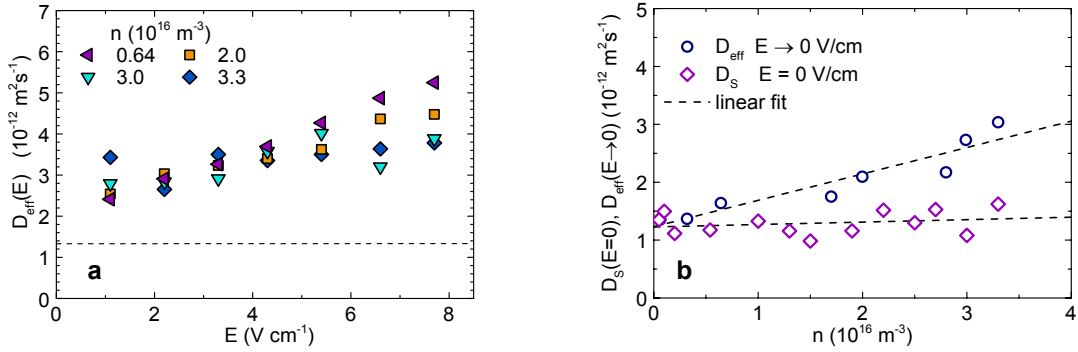

**Fig. 11:** (color online) Diffusive properties of PS310 with and without applied field. a) Field dependence of the effective diffusion coefficients obtained from fits to noise and MS corrected Doppler spectra at number densities indicated. b) Comparison of the field free self-diffusion coefficient (diamonds) to the zero field extrapolated effective diffusion coefficients (circles). At any finite concentration, the extrapolated zero field effective diffusion coefficients are appreciably larger than the self-diffusion coefficients obtained in the field free case.

PnBAPS118 crystallizes for $n > 0.2\times10^{18}$ m$^{-3}$ under deionized conditions. Application of shearing flow generates an isotropic meta-stable melt of fluid order, from which the samples readily re-crystallize after cessation of the shear flow. MS correction becomes necessary for $n > 5\times10^{19}$ m$^{-3}$. Incoherent scattering dominates over coherent scattering at low angles. The forward scattering characteristic is less pronounced and additional Bragg reflections are present at larger angles (c. f. Fig. S5). This allows studying the zero field self-diffusion coefficient during and shortly after the crystallization. In general, $D_S$ decreases with increasing $n$ and with time. For the early measurements we averaged over many repeated shear-melting – re-crystallization cycles. This ensured an acceptable quality of the spectra for each time window. In Fig. 12, we display the observed trends obtained for two series of crystallization/coarsening experiments at $n = 0.5\times10^{18}$ m$^{-3}$ and $n = 2.5\times10^{18}$ m$^{-3}$. At all times, diffusion is way below $D_0$ and notably slower in the case of the more concentrated and hence more strongly interacting sample. With increasing coarsening time, the data approach plateau values. This occurs faster for the less concentrated sample.

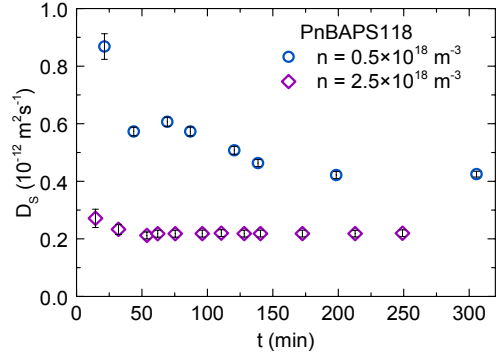

**Fig. 12:** (color online) Temporal development of the zero field self-diffusion coefficient for two series of crystallization/coarsening experiments at $n = 0.5\times10^{18}$ m$^{-3}$ (circles) and $n = 2.5\times10^{18}$ m$^{-3}$ (diamonds).

**DISCUSSION**

We performed super-heterodyne Doppler velocimetry on turbid suspensions of charged particles in the presence and absence of electric fields. Standard super-heterodyne theory outlined above could be used to evaluate the electro-kinetic spectra of PS310 in the range of $8\times10^{14}$ m$^{-3}$ ≤ $n$ ≤ $5\times10^{15}$ m$^{-3}$. For larger concentrations, an additional spectral feature appeared, not considered in SS theory. We introduced and successfully tested an empirical isolation and correction scheme for this spectral component. The isolated contribution was found to be present only in the case of sample illumination by two mutually frequency shifted beams. This identifies it as genuinely super-heterodyne. Further it has a spectral shape clearly distinct from the SS super-heterodyne signal in the presence of electro-kinetic flows (as described by Eq. (1) to (5) and (S1)). This requires it to arise from multiple scattering events of light originating from the illumination beam.

The MS contribution observed in our super-heterodyne experiment with restricted detection volume displays some remarkably characteristic features: Most importantly, it can be excellently modelled by a simple Lorentzian of well-defined finite width $w_{MS}$. For both increasing particle number density and increasing shear, this simple line shape is retained. Its width increases linearly with both $n$ and applied field strength. But even at the largest number densities, it appears to be only two orders of magnitude wider than that the SS spectral component. The MS integrated spectral power increases with increasing $n$ and depends linearly on the reference beam intensity. Under flow, the center of mass of the MS Lorentzian appears to be shifted, but no electro-kinetic information can be obtained in a reliable way. Most of these observations were not expected from our experience with MS in standard homodyne dynamic light scattering. They may, however, be partially rationalized from recalling the conceptual differences in homodyne and super-heterodyne techniques.

Homodyning detects mutual phase shifts between photons scattered under in-phase illumination. Any MS event with its ill-defined phase due to the arbitrary location of the intermediate scatterer destroys this phase coherence. It thus leads to a fast decorrelation of $C_{shet}(\mathbf{q},\tau)$ translating in into a broad (and ill-defined) homodyne peak at $\omega \approx 0$ (c. f. Eqn. (4)). By contrast, super-heterodyning signals are independent of the mutual distance of the scatterers and the phase difference of light simultaneously scattered from different particles. They only depend on the instantaneous motion of these particles during the scattering event. Taking this motion to be composed of a drift velocity and diffusion, one obtains a frequency shifted Lorentzian in the case of SS. Its-amplitude is limited by attenuation of the reference beam and attenuation of the SS light by additional scattering events. In the case of $N_{MS}$ scattering events, one observes a convolution of $N_{MS}$ Lorentzians (corresponding to the multiplication of $N_{MS}$ $C_{shet}(\mathbf{q},\tau)$ in time domain). This again yields a Lorentzian with width [17]

$$w_{MS} = N_{MS} D_{eff} q_{MS}^2 \qquad (10)$$

where $q_{MS}$ is the average scattering vector for multiple scattering. The signal strength of the MS contribution scales with $I_r <I_N>$ and thus is limited by reference beam attenuation. It is further limited by attenuation of the scattered light on its path through the sample. Most importantly, however, it is restricted by the requirements, that any detectable scattering event has to be located inside the detection volume and the direction of the photon after the last scattering event has to be parallel to the reference beam direction. In fact, most of the MS light goes undetected. This can be rationalized in a qualitative way by considering a typical scattering event at moderate turbidity (c. f. Figs. S11 in the supplementary material. Example photon paths and corresponding **q** are shown in Fig. S13).

Eqn. (10) can be used to estimate the average number of MS events, if some further assumptions about $q_{MS}$ are made. For isotropic or weakly ordered suspensions, there is only little restriction to the scattering directions from S(q), but for large particles there is a strong bias for forward scattering due to their form factor P(q). For PS310, we assume the average scattering vectors for double and multiple scattering to be mainly determined by P($q$) and the detection volume restriction. For PnBAPS118 they will be determined by the product P(q) S(q) and the same geometrical restriction. For simplicity, the average scattering vector, $q_{MS}$, is further assumed to be independent of number density. In the limit of vanishing $n$ the number of scattering events responsible for the MS-contribution is $N_{MS} = 2$, i. e. only double scattering (DS) is observed. We extrapolated the $w_{MS}$ data for PS310 shown in Fig. 7a to $n = 0$ to obtain the limiting width for DS and an average double-scattering vector $q_{DS} = (w/2D)^{1/2} = 4.2 \mu m^{-1}$. This corresponds to an average scattering angle of 18.2° which is much larger than the SS-angle but smaller than the angle at half width of P($\Theta$) (c. f. Fig. S5a). We finally assume that on average, the same geometric restrictions apply to every scattering event, irrespective of the scatterer location. Setting $q_{DS} = q_{MS}$, we calculate the average number of scattering events as $N_{MS}(n) = w_{MS}(n) / q_{MS}^2$ and plot the result in Fig. 13a. $N_{MS}$ increases roughly linearly with increasing $n$ and reaches a value of about 5 at $n = 4 \times 10^{16}\,m^{-3}$.

The field strength dependence of the MS central frequency is approximately linear at each $n$. At low $n$ where double scattering prevails, it is weaker than but still close to that of the SS central frequency corresponding to the particle velocity **v** in the mid-cell x-z-plane. Dependence appears to become weaker for increased $n$, where $N_{MS}$ gets large. This is shown in Fig. 13b for four selected number densities. We attribute this effect to an averaging of **q**·**v** over all successive scattering events. Here, **v** follows from the applied field direction and the average velocity, i. e. $v_{ep}$, and **q** follows from the direction of the incoming photon and P(q)S(q), as well as from the geometric restrictions. For $n = 6.5 \times 10^{15}\,m^{-3}$ DS prevails and the most probable q are those resulting from forward scattering as discussed in Fig. S12 of the supplementary materials. Thus two small angle scattering events add up to yield a total scattering angle $\Theta_S$. In turn, the total **q**$_2$ stays close to **q**. For larger $n$, however, **q** is successively averaged out to zero, and **q**·**v** drops.

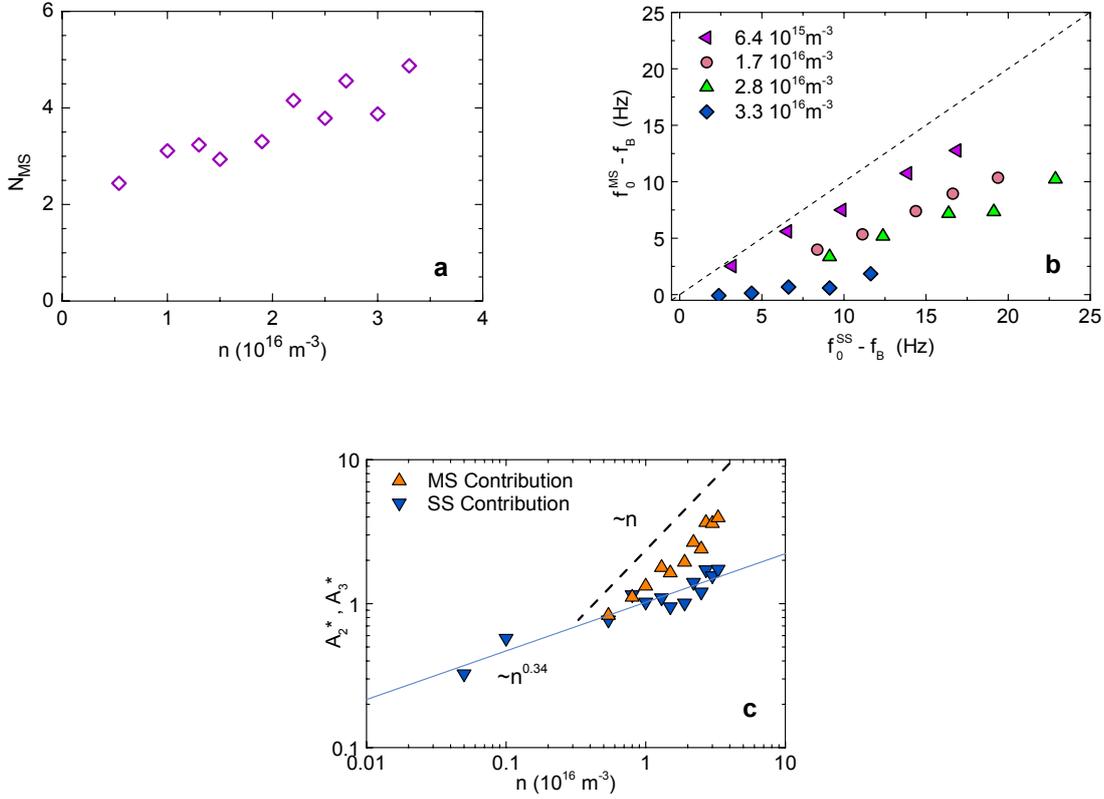

**Fig. 13**: (color online) a) Average number of scattering events contributing to the MS signal. b) Central frequencies of the MS Lorentzian plotted versus the centre of mass frequencies of the SS contribution for different particle densities as indicated. In each data set the strength of the applied field was increased in steps from left to right. The dotted line is a guide to the eye denoting $f_{0.MS} = f_{0.SS}$. c) Log-log plot of the data in Fig. 8b showing the noise normalized integrated spectral powers $A_i^* = A_i / A_1$ for SS and MS. Blue line: least square fit to the SS data returning $A_2^* \propto n^\alpha$ with $\alpha = 0.337$; dashed line: guide to the eye with $A_3^* \propto n$.

Another interesting feature is the dependence of the integrated spectral power on number density. In Fig. 13c we plot the data of Fig. 8a in a log-log fashion to check for power law behavior. Compared to the expectations arising from the pre-factor of heterodyne scattering in Eqn. (4), both the SS and the MS contribution show a less pronounced than expected dependence. Instead of a direct proportionality, the fit of the $A_2^*$ data over more than a decade in $n$ returns a power law of exponent $\alpha = 0.337$. The MS data appear compatible with $\alpha = 1$. This differs from the expectations given for SS in Eqn. (3) and (4) from which a linear scaling of $A_2$ with $n$ is expected. Such a scaling indeed has been observed previously in well-ordered, single scattering systems of small particles (diameter 70nm) where their coherent scattering cross section was very small and the dominant incoherent scattering not prone to attenuation [19]. Our observation also deviates from the simple expectation for attenuation-free MS for which $A_3^* \propto I_r <I_{MS}(q)> = I_r <I_1(q)> <I_2(q)> <I_3(q)> \ldots \propto n^{N_{MS}} P(q) S(q)$.

Several possibilities for a sub-linear increase of $A_2^*$ and a liner increase in $A_3^*$ may in principle be considered. In the present case going from non-interacting to weakly structured suspensions, we cannot exclude some influence of structure formation. Even under the present conditions of strong incoherent scattering, this may result in a decrease of S(q) at small angles and enter into $A_2^*$ and $A_3^*$. Further, we may expect the influence of attenuation of the reference beam entering through the pre-factors in Eqn. 4 for every successive scattering event. However, at the same time one should also account for attenuation of the illumination beam on its way through the sample, the loss of detectable scattered light by additional scattering events, and the generation of detectable light by scattering from paths not parallel to the reference beam direction. A full theoretical description explicitly including all these gain and loss terms for single scattering and every following generation of scattering events is clearly beyond the scope of this paper. At present we therefore cannot explain this behavior and regard it an interesting observation occurring in the case of attenuation by multiple scattering as well as a constraint for theoretical descriptions to be developed in future.

The next point to discuss is the usefulness of the here proposed and tested scheme in other turbid systems and for other physical properties. Standard dynamic light scattering can be applied for samples of transmittance T ≈ 0.9-0.95 [7,55]. Our experiments were performed on particles showing dominant incoherent scattering with more or less pronounced forward scattering characteristics. This allowed neglecting the influences of structure formation on electro-kinetic and diffusion spectra. MS became discernible for T ≈ 0.9 and our scheme was successfully tested to T≥0.4 but may in principle be applied up to limit of the Beer Lambert law, which for PS310 is at $n = 1 \times 10^{17}$. Applications may include rheological measurements or measurements of active particles, both up to high particle concentrations and strong mutual interaction strengths.

We profited from super-heterodyning and a strong restriction of the detection volume, such that we were not concerned with homodyne phase decorrelation and most of the MS light produced went undetected. Our method appears to be very flexible and our correction scheme should be readily applicable also many other cases without the need for instrumentally demanding cross correlation or mode selection schemes. We measured particle velocities and self-diffusion in ergodic and non-ergodic samples. For the latter, standard homodyne experiments require additional efforts to obtain a correct averaging in order to apply the Siegert relation and calculate $g_E(q,t)$ from $g_I(q,t)$. [56]. For the presently studied poly-crystalline samples, ensemble averaging is provided by the large illumination volume which exceeds that observed in a typical dynamic light scattering experiment by about two orders of magnitude. Moreover, super-heterodyning directly accesses $g_E(q,t)$ instead of $g_I(q,t)$. This makes it very interesting to test our approach also in other samples and at larger angles. The present experiments were performed only at a single angle without yet exploiting the possibility of q-dependent measurements. Preliminary experiments show that we can use the correction scheme also for angles up to some 15°. It remains, however to be tested, whether it can still be applied in a standard light scattering goniometer when illumination beam and reference beam are crossed at 90°.

In priciple, we do not see any principle objection to applying our correction scheme also to samples with dominant coherent scattering. However, a few issues may arise. One is discrimination of SS and MS contributions. Since super-heterodyning does not record phase de-coherence, but successive Doppler broadening events, no strong separation of time scales is observed. With $w_{MS} = N_{MS}Dq_{MS}^2$ and $D$ being the same for MS and SS, signal discrimination may become difficult where $q \approx q_{MS}$, For diffusion measurements close to the maximum of S(q), it may be useful to employ minute plug-flows, which do not interfere with structure (c.f. Fig. 7c) and perform discrimination *via* the location of the spectral maxima. Alternatively, one may increase the particle scattering cross section (e.g. by refractive index mismatching) to increase $N_{MS}$. A second issue may be the discrimination of relaxation processes occurring on different time scales [57]. For glassy samples with well separated short and long-time dynamics a superposition of Lorentzians is expected for both SS and MS. A convolution of a Lorentzian with a distribution of line widths can be expected. It remains to be tested, how these can be discriminated. In general, however, measurements of stretched exponentials or power-law decays or other complex temporal behavior, which are easily feasible in standard dynamic light scattering [58], appear to be difficult in frequency domain due to the *a priori* unknown spectral shape. A workaround may possibly be provided by Fourier-back-transformation.

In the present study, our correction scheme was applied to obtain different physical properties from the isolated SS-spectra. It was utilized i) to investigate the spectral broadening under electro-kinetically driven solvent flow flow and particle drift; ii) to quantify Brownian motion in the field free case in differently ordered samples; and iii) to study the drift motion in an electric field. Concerning diffusion measurements with applied electric field, our data support earlier findings of a linear dependence of $D_{eff}(E)$ on field strength respectively the average shear rate $\gamma = dv / dx$. This demonstrates the possibility to apply our approach also to sheared systems, not easily studied by standard dynamic light scattering. Moreover, we could for the first time observe a pronounced $n$ dependence of the field dependent broadening. This seems to favor field induced velocity fluctuations [12] or a polydispersity of the electro-phoretic mobility [49] over Taylor dispersion [41] accounting for effects of shear. The mobility polydispersity scales directly with the size polydispersity [27] and therefore should follow the n-dependent mobility. Velocity fluctuations are expected to be most pronounced when the structural correlation length is on the order of the interparticle distance, i. e. at intermediate $n$, where the structure is still weak. By contrast, shear in our experiments depends on the electro-osmotic velocity (c. F. Eqn. S1), which shows only very little dependence on $n$.

For solidifying PnBAPS118, the field free self-diffusion coefficient was observed to decrease to a limiting value. The more dilute sample showed the slower temporal evolution. This is in line with previous experiments using Forced Rayleigh scattering or microscopy [59, 60]. We therefore believe, that such experiments contain interesting information about grain boundary dynamics. Future experiments will complement existing structural studies on coarsening samples [61, 62, 63, 64].

For the electro-kinetic measurements on PS310 our approach facilitated an extension of the accessible volume fraction range by nearly one order of magnitude (c. f. Fig. 12). We found that under low salt conditions, the electro-phoretic mobility, $\mu_e$ first increases, then displays a plateau and finally decreases again. The initial ascent is in line with earlier observations on extremely low mobilities for isolated particles and on particles observed at low number densities [24, 25, 26, 27]. The plateau and final descent are compatible with previous observations made on particles at large number density under conditions of counter-ion dominance [27, 28, 29, 30]. The previous data were gathered for particles of different surface chemistry and size, at fully deionized conditions and in the presence of salt. In view of the present findings we propose that each has been covering different parts of a universal curve.

A rigorous quantitative test, however, would in addition require a precise control of the salt concentration. In the present study, we worked under $CO_2$-saturation without conductometric control in both transmission and electro-kinetic experiments to test the correction scheme at known turbidities. Any residual uncertainty of the electrolyte content, however, can have large effects on mobility. Measurements with precise adjustment of different low salt concentrations are under way.

**CONCLUSION**

We have introduced a simple yet powerful empirical MS-correction scheme for super-heterodyne light scattering in frequency domain. We demonstrated its scope and limits by systematic LDV measurements on different charged particle suspensions. We discussed its versatility and limits and indicated a variety of possible applications beyond electro-kinetics. We anticipate that this approach will allow for a number of interesting investigations on the dynamic properties of turbid suspensions at moderate to a very low transmittance. A full theoretical description of the MS contribution to describe its interesting characteristics is highly desired.

**SUPPLEMENTARY MATERIAL**

See supplementary material for additional details concerning the experimental setup, particle flow profile and multiple scattering contribution.

**ACKNOWLEDGEMENTS**

We gratefully acknowledge many fruitful and in the best sense, controversial discussions on electro-kinetics with our theoretical colleagues M. Heinen, A. Delgado, F. Carrique, E. Ruiz-Reina, R. Roa, V. Lobaskin, B. Dünweg, and C. Holm who strongly motivated our electro-kinetic study and therefore also the technical improvements made. We are indebted to G. Nägele for his theoretical support in issues of heterodyning, to E. Bartsch for critical discussions of multiple scattering issues. We thank S. Heidt and R. Sreij for the PI-determination and R. Dekker for providing the S(q) of PnBAPS118. Financial support of the DFG (SPP 1726 and Pa459/18) as well as of Interne Forschungsförderung, JGU is gratefully acknowledged. L.M.M. further gratefully acknowledges financial support by the DAAD IAESTE program.

# Supplementary material

**Technical details of the super-heterodyne integral low angle Doppppler velocimetry experiment**

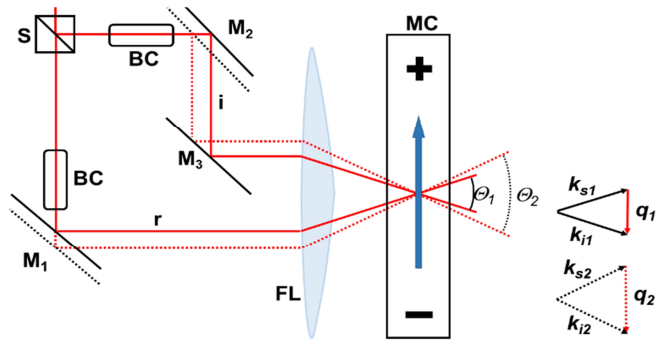

**Fig. S1:** (color online) Sketch of the experimental scattering geometry with the optical paths of reference and illuminating beams, r and I as indicated. To the right, we show the resulting scattering vector **q** for different positions of the coupled mirrors $M_1$ and $M_2$ and corresponding beams paths as indicated by the solid and dashed red lines. For details see text. The direction of particle motion opposite to that of the applied electric field is indicated by the blue arrow. Beam propagation is drawn in air neglecting refraction effects in the cell.

The beam of a He-Ne laser of circular frequency $\omega_0 \approx 10^{15}$ Hz is split into two beams, a reference beam (r) and an illuminating beam (i), using a beam splitter (S). Each beam passes through a Bragg-cell (BC, Model 3080-125, Crystal Technology Inc., US) with a frequency shifter DFD 80, (APE GmbH, Germany) and is frequency-shifted by the circular frequencies $\omega_i / 2\pi = 80.002$ MHz and $\omega_r/2\pi = 80.000$ MHz, with the larger frequency shift realized for the illuminating beam (upper beam in Fig. 1). This results in a positive relative Bragg-shift, $\Delta\omega_B = \omega_i - \omega_r$ in the circular frequency corresponding to $\Delta f_B = \Delta\omega_B / 2\pi = (2000\pm0.05)$ Hz. The beams are redirected into the sample cell (MC) by a set of mirrors ($M_1$ to $M_3$) and a lens (FL). There they cross in the suspension under an angle $\Theta = (5 - 28)$°. (Note that in Fig. S1, we show the beam propagation in air.) An improvement as compared to previous set-ups was adapted from references [1] and [2]. In the present setup we can, in principle, vary $\Theta$ continuously simply by shifting the mechanically coupled mirrors $M_1$ and $M_2$. In Fig. S1, a downward shift to the dotted positions increases the scattering vector **q**.

On the detection side (Fig. 1 in the main text), a lens, an aperture and a polarizer, to assure V/V detection, as well as grin lens mounted on an optical fiber are placed on the reference beam optical axis. For our so-called integral measurements, light scattered off the particles is received by the lens and focused in its focal plane. The detection optics arrangement defines the maximal extension of the cylindrical detection volume that encompasses the complete illumination beam, the scattering vector **q** and the q-resolution $\Delta q$. It collects singly scattered light emerging off the region illuminated by the illumination beam as well as multiply scattered light from within the complete detection volume. Both contributions are propagating in the direction of the reference beam. The fiber is leading to a photomultiplier used to record the intensity.

**Separation of homodyne and super-heterodyne spectral contributions in frequency space**

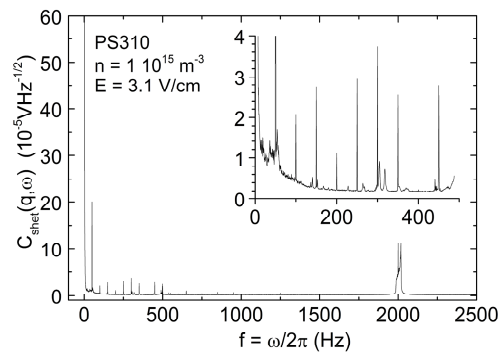

**Fig.S2**: Example of a complete positive LDV spectrum for PS310 at $n = 1 \times 10^{15}$ m$^{-3}$. One notes a $\delta$-term at zero frequency, superimposed with the homodyne contribution and positively Bragg-shifted super-heterodyne part. For this sample no MS contribution is detectable in the region of super-heterodyne scattering. Further note the absence of the noise in the super-heterodyne range around 2 kHz. Inset: homodyne part of the spectrum. Note that the spectral width of the homodyne peak centered at zero frequency is on the order of 100Hz and much larger than the broadening observed for the heterodyne signal.

According to Eq. (4), the complete Doppler signal is composed of three terms. The desired information about drift velocities and diffusional properties is contained in the frequency-shifted super-heterodyne term. The spectral range of the homodyne term is in addition contaminated with contributions of electronic noise and vibrations. Even in the absence of shear and MS, it cannot be evaluated without considerable extra correction effort. An example for a complete positive part of the spectrum is displayed in Fig S2. The super-heterodyne part is well separated from the homodyne term containing no contaminations except for a δ-function type contribution at $\Delta f_B$ = 2kHz. This trivial contribution is not considered by our theory and stems from the beat between parasitic scattering at the walls by glass inhomogeneities, scratches or adsorbed stationary particles. At the chosen low particle concentration, no MS-contribution is detectable in this frequency range. The homodyne part however shows a considerable broadening of the Lorentzian underlying the noise and power grid peaks, which is on the order of 100Hz. This cannot be explained by single scattering according to Eqn. (4) and indicates a significant contamination by MS present in the homodyne signal.

**Flow profiles and Doppler shift distributions**

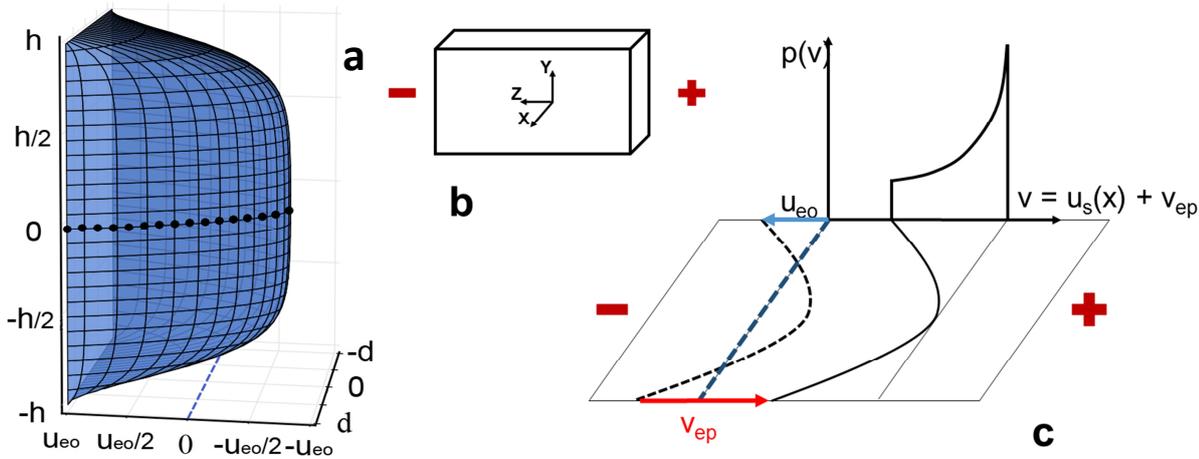

**Fig. S3:** (color online) solvent and particle flow. a) Solvent velocity profile $\mathbf{u}_s(x,y)$ inside a cell of $h/d = 5$ in units of the electro-osmotic solvent velocity at the walls $\mathbf{u}(x = |d|, y) = \mathbf{u}(x, y= |h|) = \mathbf{u}_{eo} = \mu_{eo}\mathbf{E}$, where $\mu_{eo}$ is the electro-osmotic mobility. Note the pronounced central backflow at $x \approx 0$ which is nearly independent of $y$. The black dotted line denotes the cell's mid-plane. b) Sketch of the closed measuring cell with coordinate system and electrodes indicated. Thickness of the cell in $x$-direction is $2d$, height is $2h$, $z$-axis points in field direction. c) Particle velocity distribution in the mid-cell $x$-$z$-plane at $y = h = 0$ for superimposing solvent flow and electro-phoretic motion. The blue dashed line denotes zero velocity. Electrode positions are indicated. Negatively charged colloids move to the right with a negative velocity $\mathbf{v}_{ep} = \mu_{ep}\mathbf{E}$ and a negatively charged cell wall induces a negative solvent velocity in the center. Both effects superimpose and the resulting particle velocity distribution, $p(v)$ is shifted rightward away from the origin.

In the absence of flow, the normalized particle velocity distribution, $p(v)$ and the corresponding distribution of Doppler frequencies, $p(\omega_D)$ are $\delta$-functions centered at $v = 0$ m s$^{-1}$ and $\omega_D = 0$ Hz, respectively. The super-heterodyne Lorentzians are located at $\pm\omega_B$. Application of a homogeneous electric field **E** to the suspension leads to an electro-phoretic motion of the charged particles with respect to the solvent and an electro-osmotically induced solvent flow. Our measurements are performed in a closed cell with negatively charged walls. Application of an electric field therefore leads to an electro-osmotic flow along the cell walls and a counter-propagating backflow of the incompressible solvent in the cell center. In Fig. S3a, we sketch the resulting solvent flow profile $\mathbf{u}_S(x,y)$ for a cell of height $2h$ and depth $2d$ in units of the electro-osmotic velocity at the wall, $\mathbf{u}_{eo} = \mu_{eo}\mathbf{E}$, where $\mu_{eo}$ is the electro-osmotic mobility. The lab coordinate system is shown in Fig. S3b. Here, $x$-, $y$-, and $z$-directions point in the direction of the optical axis or cell depth direction, the cell height direction and the direction of the field, respectively. The origin is taken to be at the cell center.

The flow profile was calculated according to [3] using the boundary condition of no net solvent flow. The particle velocity then results as the sum of particle electro-phoretic velocity and solvent profile $\mathbf{v} = \mathbf{v}_e + \mathbf{u}_S(x, y, \mathbf{E})$. Here, $\mathbf{v}_{ep} = \mu_{ep}\mathbf{E}$ is the electro-phoretic velocity relative to the solvent depending on its electro-phoretic particle mobility $\mu_{ep}$ and the electric field **E**. In our experiments we measure the velocity distribution at mid-plane cell height ($y = 0$). Under this conditions the particle flow profile, $\mathbf{v}(x, y = 0)$ can be analytically approximated as [4]:

$$\mathbf{v}(x, y=0) = \mu_{ep}\mathbf{E} + \mu_{eo}\mathbf{E}\left[1 - 3\left(\frac{1 - \frac{x^2}{d^2}}{2 - \frac{384}{\pi^5 K}}\right)\right] \tag{S1}$$

Here, $K = h / d$ is the ratio of cell height, $h$, to cell depth, $d$. The resulting mid-plane flow-profile is sketched in Fig. S3c for $K = 5$. It is parabola-like and the corresponding $p(v)$ is asymmetric with a pronounced maximum at a non-zero center velocity [5].

**Optical characterization of samples**

The region, from which multiply scattered light emerges, can be visualized in images taken from above the cell illuminated by a vertically polarized laser beam. Due to the Hertzian scattering characteristic, these images contain singly scattered light only via the contribution of depolarized scattering. Light scattered coherently within the detection plane is still polarized. Multiple scattering events leaving the x-z-plane, however, depolarize the photons, such that multiply scattered photons with their last scattering direction towards the camera are imaged as reddish glow. This glow is clearly seen in Fig. S4 for two concentrations: a cone shaped reddish region with its broader and at the same time brighter side located towards the detector as well as a broadened illumination beam.

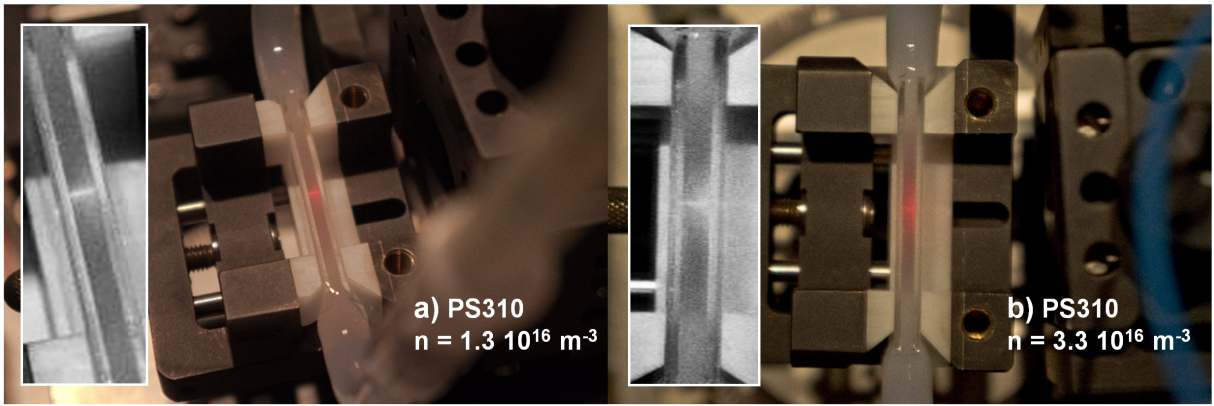

**Fig. S4**: (color online) images of a 2mm sample cell with PS310 dispersion at concentrations as indicated and crossed by illuminating beam only. The illumination beam enters the cell on the left side. The images demonstrate the distribution of secondary and higher order illumination by photons scattered off the illuminating beam. Viewed from above, the beam appears broadened and a diffuse, several mm wide MS scattering cone is identified from its reddish glow (or lighter gray appearance in the black-and-white close-ups.). By restricting the detection volume to a sheet-shaped region very close to the reference beam most of the scattering resulting from secondary and higher order illumination is excluded.

The cone shape of the glowing region is expected considering the strong forward scattering characteristic of PS310. The latter is illustrated in Fig. S5a for PS310 in terms of the angular intensity distribution derived from a calculated P(q). With increasing $n$, in addition, a static structure may evolve and the angular intensity distribution is slightly modified, shifting the maximum intensity out to a finite angle. For the $n$ shown in Fig. S4 a and b, these angles would be 8.7° and 12°, respectively ($\Theta_S$ = 6°). This effect, however is only weakly pronounced for PS310 at $CO_2$-saturated conditions The combined effect of forward scattering and structure formation is much more pronounced for PnBAPS118 at $n$ = $4.7 \times 10^{18}$m$^{-3}$ and under fully deionized conditions. This is shown in Fig. S5b using an S($\Theta$) measured independently in a static light scattering experiment

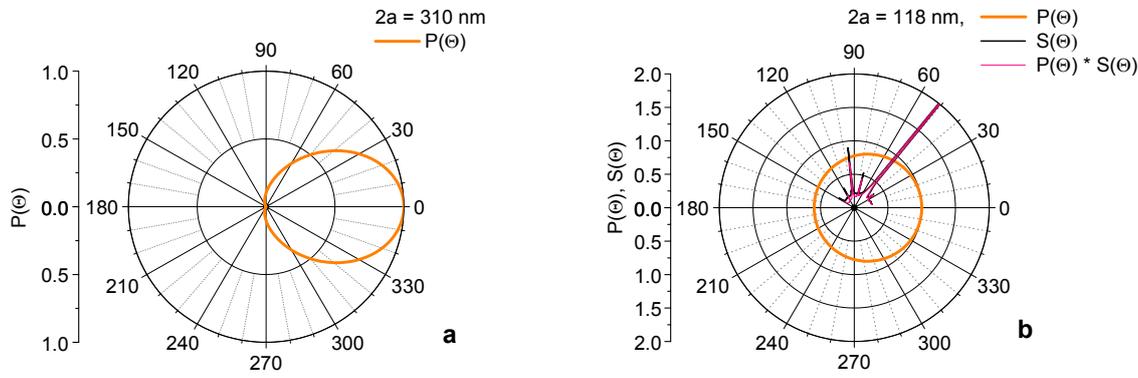

**Fig. S5**: (color online) Distribution of singly scattered light in dependence on scattering angle. a) calculated form-factor, P(Θ), of PS310 being a very strong forward scatterer. b) Form factor, P(Θ), experimental static structure factor S(Θ) as measured in independent experiments and product P(Θ)×S(Θ) contributing together with $b^2(0)$ and $nV$ to Eqn. (3). For comparison, the SS scattering angle in the present electro-kinetic and diffusion experiments is $\Theta_S = 6°$

With or without structure formation, the light cone formed for single scattering gradually fans out with every additional scattering event. The strong cone broadening observed in Fig. S2 therefore indicates the presence of several scattering events in the path of the camera-recorded photons even at $n = 1.3 \cdot 10^{16}$ m$^{-3}$.

In addition to the MS fan, the vertically polarized illumination beam is visible as narrow, slightly blurred high intensity region. We attribute this to the presence of incoherent scattering originating from the size polydispersity of PS310. In this case the probability distribution for the scattered intensity is given by $I(q) \propto P(q)S_M(q) = P(q)$. Furthermore, the probability for photons scattered in off-beam direction at their first scattering event to return to the beam location for their next scattering event is negligibly small. Thus, the beam broadening visible in Fig. S4 is attributed to an initial depolarizing incoherent scattering event followed by additional forward scattering events into the direction of the camera.

**Sample conditioning and charge characterization**

Sample preparation used procedures are described in detail elsewhere [6, 7, 8]. Pre-cleaned stock suspensions were diluted with milli-Q grade water. Samples were loaded into a thermostatted peristaltic conditioning circuit. The circuit contains an ion exchange column filled with mixed bed ion exchange resin (Amberjet, Carl Roth GmbH & Co. KG, Germany), a reservoir to add water or further stock suspension, a conductivity experiment (electrodes LR325/01 and LR325/001; bridge LF340i, WTW, Germany) and a cell for static light scattering, all connected via Teflon® tubings. The suspension is peristaltically cycled and the drop of conductivity with progressing deionization is noted. With pure water, the set-up suffices to reach conductivities somewhat below 60 nScm$^{-1}$ after some ten to twenty minutes. From the larger final conductivity reached with particles added, we estimate the residual ion concentration in the thoroughly deionized state to be on the order of c ≤ 1×10$^{-6}$ mol/l [7]. The suspension static structure factor was measured in a home-built multi-purpose light scattering instrument [9] and evaluated for the particle number density $n$. The number density dependence of the conductivity of PS310 is shown in Fig. S6. From the slope we derive an effective number of freely moving H$^+$ - counter-ions, $Z_{eff}$ = 5260±100 using the model of independent ion conductivity [7, 10]. Within charge renormalization theory, this corresponds to saturation value of $A$=8.01k$_B$T [11]. The effective charge for PnBAPS118 is $Z_{eff}$ = 647±18.

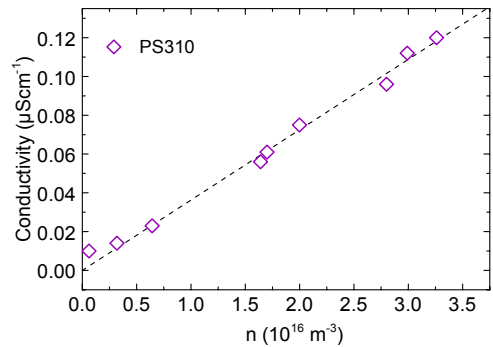

**Fig. S6:** (color online) $n$ dependence of the low frequency AC conductivity of deionized PS310. The dashed line shows a fit of the theoretical expectation for this quantity within the model of independent ion conductivity [7, 10]

For PnBAPS118, the preparation circuit was then disconnected from the static light scattering cell and connected to the electro-phoretic cell. This way we could easily alter the particle concentration but maintain a thorough deionization under conductometric control. After conditioning, the measuring chamber can be isolated from the remaining circuit by electromagnetic valves to avoid residual flows and minimize contamination with airborne $CO_2$. For number densities above $n$ = 0.2 μm$^{-3}$, PnBAPS118 then readily crystallizes in a body centered cubic structure. In the crystalline samples, a minute plug flow can be applied, when needed, by opening the valves and moving the suspension with a syringe pump at constant flow rate.

**Additional characterization of the MS signal**

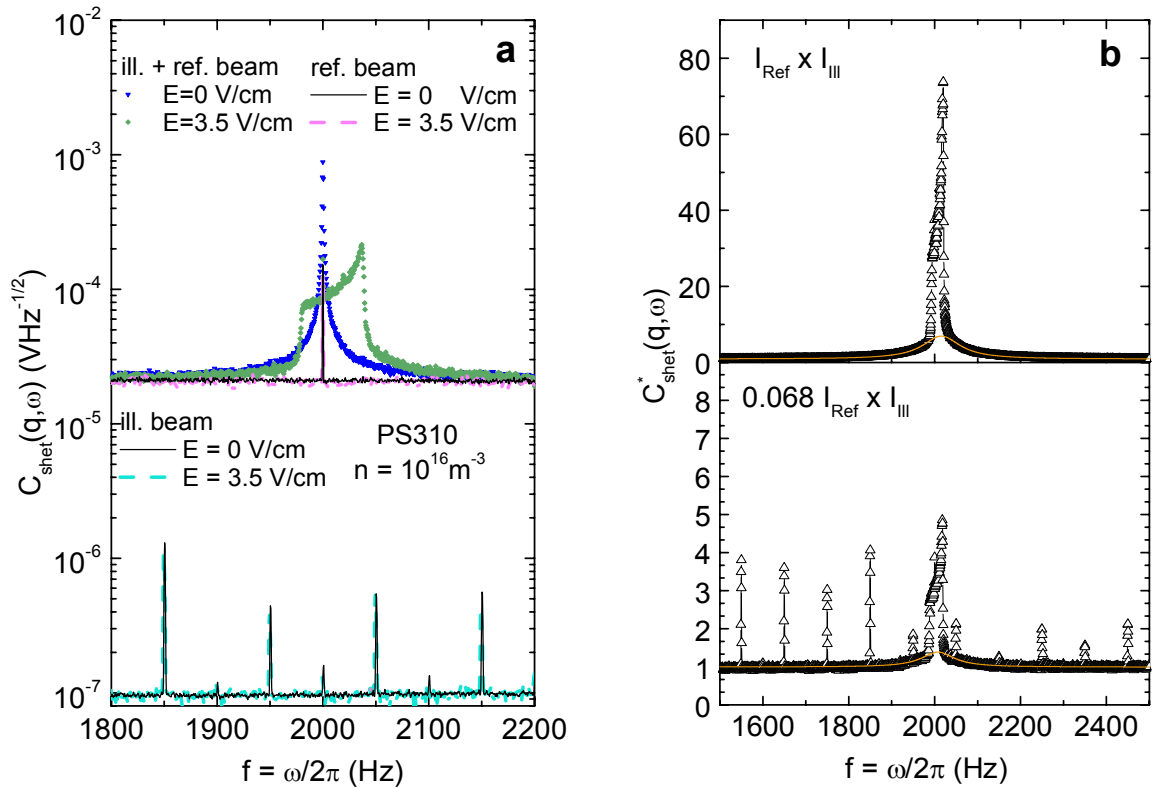

**Fig.S7:** (color online) Characterization of the MS contribution. a) Absolute super-heterodyne signals of PS310 at $n = 1.0 \times 10^{16}$ m$^{-3}$ measured at different illuminating conditions and concentrating on the super-heterodyne spectral range close to $\omega_B$. All data plotted on a single scale. Spectra composed from reference and illuminating beams with $E = 0$ Vcm$^{-1}$ (triangles) and $E = 3.5$ Vcm$^{-1}$ (squares). Reference beam only with $E = 0$ Vcm$^{-1}$ (solid blue line) and $E = 3.5$ Vcm$^{-1}$ (dashed magenta line). Illuminating beam with $E = 0$ Vcm$^{-1}$ (solid black line) and $E = 3.5$ Vcm$^{-1}$ (dashed cyan line). b) Noise normalized spectra (triangles) with MS contribution fitted (orange line). Top: spectrum recorded at standard illumination conditions. Bottom: reference beam dimmed to 6.847% of its original value. Except for the power grid noise, all spectral components (SS, MS and background noise) scale linearly with the reference beam intensity.

In Fig.S7a, we compare the absolute values for $C_{shet}(q,\omega)$ for different combinations of beams detected in the same frequency range centered about $\omega_B$. Using both illuminating and reference beams corresponds to the standard measuring configuration. We compare this signal to the signal for reference beam only and illuminating beam only. With both beams present the characteristic SS-signal shape and the additional MS contribution are superimposed on a noise background. This contribution remains at a value of $2 \cdot 10^{-5}$ V Hz$^{-1/2}$, if the illuminating beam is turned off. With the illuminating beam only no signal is detected in the super-heterodyne frequency range, except higher multiples of the 50Hz mains frequency and a low background noise at a value of $8 \cdot 10^{-8}$ VHz$^{-1/2}$. Since only the beating of the reference beam with scattered light from the illuminating beam leads to a signal in the super-heterodyne frequency region (See Eqn. (5)), both SS and additional broad background contributions must originate from such a combination. As the SS contribution can be clearly discriminated by its spectral shape obeying Eqn. (5) with (S1), the remaining additional signal has to come from MS. Furthermore, our comparison shows that the frequency independent noise background is stemming from the reference beam only.

All signal components, except the electronic noise visible as multiples of the electric mains frequency, scale in their power with the reference beam intensity. This is shown in Fig. S7b, where we decrease the reference beam intensity by placing two neutral filters of D0.7 and D0.4 in the beam path. Accounting further for a 4% reflection at each additional glass/air interface, the total optical density of this element is D1.164 corresponding to a 6.847% decrease in the reference beam intensity. The ratio of the integrated intensities of SS, MS and noise contributions stayed constant and the integrated spectral power decreased to 6.7% of the original one.

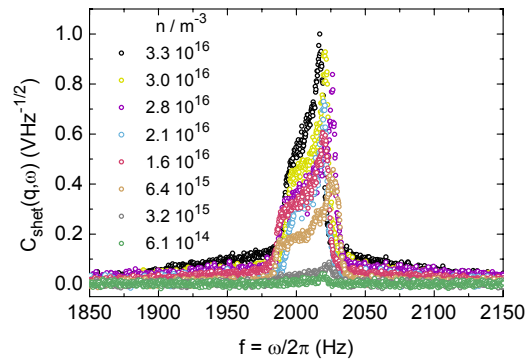

**Fig. S8**: (color online) Evolution of the super-heterodyne spectra with change of the number densities.

This actually is a nice feature for the interpretation of other characterization measurements and future systematic experiments. Sample exchange often affords removing the cell for preparation purposes and remounting for measurement. This typically results in minute differences of cell adjustment resulting in a slight shift of the reference beam position on the grin lens. Moreover in density dependent measurements, the relative contributions of SS and MS contribution vary in a systematic way. Both effects can be seen in the series of spectra taken at different $n$ shown in Fig. S8. Fig S9 shows the absolute integrated spectral power for MS Lorentzian and noise background. Observation of the strict scaling of all spectral contributions with the reference beam intensity (c. f. Fig. S7b) then allows for a convenient normalization of spectral amplitudes using that of the noise contribution. The noise-normalized integrated spectral magnitudes of the MS Lorentzian and of the SS signal are shown in the main text in Fig. 7b.

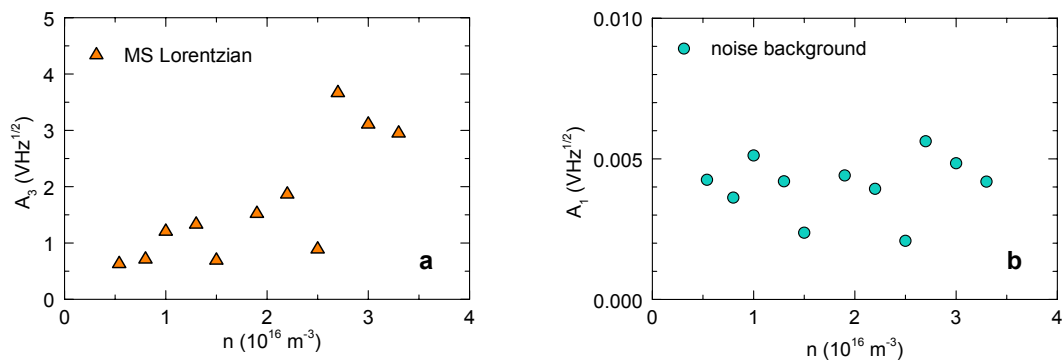

**Fig S9**: (color online) Dependence of the integrated spectral power of different signal contributions on number density for a) MS Lorentzian and b) noise background. Note the different scales. While $A_3$ increases with $n$, $A_1$ stays approximately constant. Both data sets show a considerable scatter. Note the correlation of scatter in both data sets.

**Example for MS correction under conditions of weak forward scattering**

Fig. S10 shows the Doppler spectrum of PnBAPS118 in the crystalline state obtained at high particle concentration and low electrolyte concentration. Note the different power density scale as compared to the spectra of PS310 in the main text. Also, note the sharp edges of the SS contribution as compared to the spectra of PS310 taken at the same field strength. The latter feature reflects the low diffusivity of PnBAPS118 in the crystalline state. As for PS310, the width of the MS contribution of $w \approx 10^2$ Hz is much broader than that of the SS-contribution. The relative weight of the MS contribution, however, is much smaller than for PS310. In such cases, the particle electro-kinetics can be directly obtained from the Doppler-spectra without MS correction. A reliable determination of diffusive properties, however, affords this correction further here.

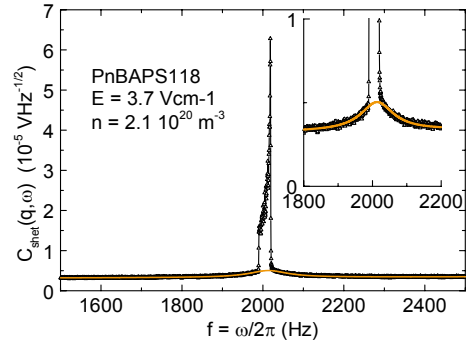

**Fig. S10**: (color online) Super-heterodyne spectrum of PnBAPS118 recorded at $E = 3.7$ Vcm$^{-1}$ with $n = 2.10 \times 10^{20}$ m$^{-3}$, $2d = 2$ mm.

## Some thoughts on $q_{MS}$- and $<I_{MS}>$ restriction

The characterization of the MS-contribution yielded a weaker than expected dependence of the integrated spectral power densities and only a moderate broadening of the MS-Lorentzian. We here give a few preliminary and qualitative suggestions on the underlying mechanisms.

Primary scattering is located on the path of the illumination beam. Two strict geometric restrictions apply to additional scattering events. They must be located inside the detection volume and the last scattering event has to fulfill $k_N = k_{ref}$ where $N$ refers to the $N$th scattering event. To illustrate this idea, we consider a situation for an ideal gas of Rayleigh scatterers, i. e. $P(q) = 1$ and $S(q) = 1$, moving with a constant velocity **v** in $z$-direction, as sketched in Fig. S11. In our experiments this corresponds to the incoherent scattering contribution of PnBAPS118. For simplicity we only draw V/V scattering restricted to the $x$-$z$ plane. Sample turbidity is characterized by a scattering length $\ell^*$ over which the transmission has dropped to 1/e along the illuminating and reference beams and also in the region illuminated by a first scattering event located on the illumination beam after a path $L_1$. Clearly, only a small number of scattering events will lead to paths $L_2$ ending in grey shaded detection volume. In the case of isotropic scattering, this restriction favors forward and backward scattering alike. For crystalline PnBAPS118, a smaller fraction of coherently scattered photons will stay in the detection volume due to the influence of $S(q)$ (c. f. Fig. S5b). In this case, scattering under Bragg angles, i.e. nearly perpendicular to the orientation of the detection volume cylinder, is favored. For PS310, the MS-characteristics are less influenced by the evolving weak structure than by the pronounced forward scattering due to its $P(q)$.

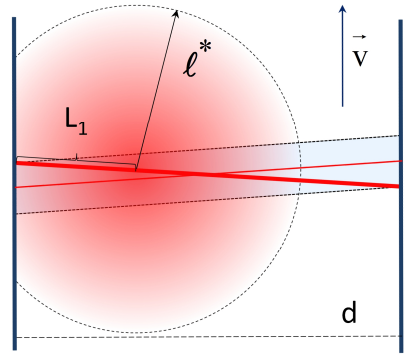

**Fig. S11**: (color online) To-scale sketch of the scattering geometry in the *x-z* plane with red parts representing illuminated regions (top view). Borders of the cell of depth $d$ = 1mm are denoted by thick vertical lines. It is traversed by both illumination beam (thick) and reference beam (thin) crossing symmetrically in the cell center, here drawn with an angle of $\Theta_s$ = 8°. The light gray shaded region is the detection volume. Particles (not drawn) are moving upward with a velocity **v**. A scattering event is sketched occurring after a path length $L_1$ inside the cell. The circle drawn about this point has a radius of one scattering length, $\ell^*$. The decrease of shading intensity corresponds to the attenuation of scattered light by following scattering events.

The second effect that has to be accounted for within each photon path is attenuation. In general, this favors the paths of shortest total length. In Fig. S12 we draw three example paths for double scattering and the corresponding **q**.

Path a) involves two backscattering events and combines two scattering vectors of large magnitude. For coherent scattering of PnBAPS118 and for PS310, $P(q)$ and $S(q)$ leads to a low probability for this path to be taken. Further, the backward scattering elongates the total path by about one third as compared to a minimal path and $L_3$ nearly equals the cell depth. This strongly increases the probability, that the photon will leave this path in an additional scattering event. Path b) has a low initial probability since an $L_1$ of length $L_1/\ell^* > 1$ will be rare. The scattering vectors in this example are favorable for coherent scattering in crystalline PnBAPS118 in directions of higher order Bragg reflections. In the chosen combination, the chance to stay in the detection volume is increased by the large value for $P(q)S(q)$ yielding a high probability to be scattered again under Bragg condition after a short path $L_2$. Chosen $q$ do not yield a high probability for PS310 scattering along this path. Path c) safely fulfils the detection volume restriction condition. Moreover its total length is close to the minimal possible length and chances for photons to survive on this path are high. For PS310 the directions taken comply very well with the strong forward scattering bias by $P(q)$. Moreover, the closeness of $\omega_0 = \mathbf{q}_{MS} \cdot \mathbf{v}_{ep}$ in MS to $\omega_0 = \mathbf{q}_{SS} \cdot \mathbf{v}_{ep}$ in SS observed at the onset of MS suggests that this path type has a very large probability under DS conditions. Also for the in-coherent scattering of PnBAPS118, paths like c) have a non-zero probability to occur (c. f. orange curve in Fig. S5b). For coherent scattering, however, the probability to take such a forward-forward path becomes vanishingly small.

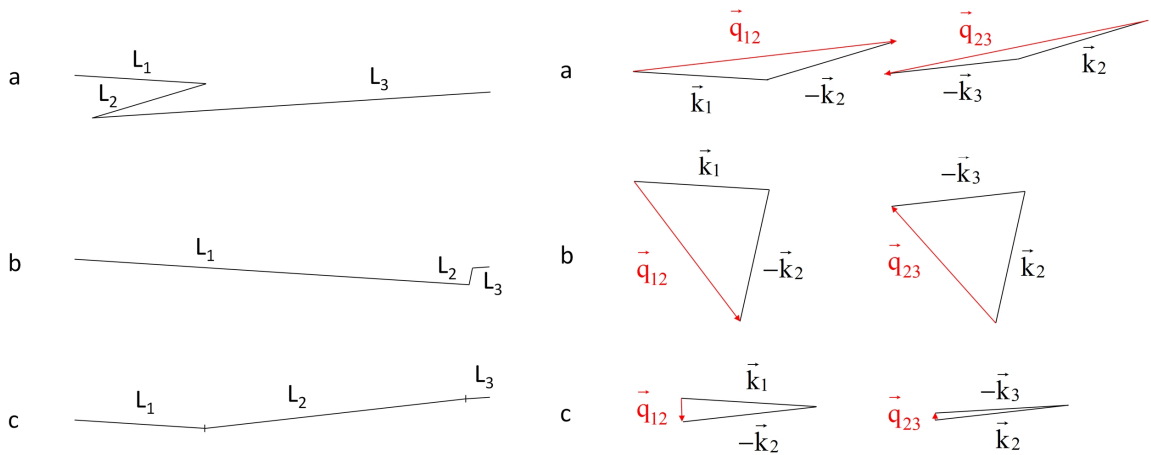

**Fig. S12**: (color online) Example paths for double scattering and corresponding $q$. a) Path involving two back scattering events; b) path involving two scattering events at intermediate angles; c) path involving two forward scattering events.

The above discussion has qualitatively shown that both attenuation and detection volume restriction will lead to a majority of detected double scattering paths zigzagging along the detection volume in forward direction. The same reasoning can also be applied for higher order MS. Hence, smaller scattering vectors are favored over long ones and the average of this this selection enters as $q_{MS}^2$ into the width of the MS-Lorentzians. We anticipate that this provides a qualitative explanation of the experimentally observed moderate broadening with $q_{MS} < q_{MAX}/2$. However, one has to keep in mind, that this describes the detectable paths only. The situation is different for the paths taken but not detected. In fact, most of the MS light is not detected but distributes throughout the complete cell and leaves in other directions. This is readily seen in the case of PS310 in the images of samples traversed by the illumination beam shown in Fig. S2.

Finally, we return to Fig. S12 and consider an increased turbidity. As $\ell^*$ shrinks, the geometric selection effect is reduced. I. e. an increasing number of paths with $q_N$ perpendicular to the orientation of the detection volume will stay within it. For PNBAPS118 this will increase the probability of detecting photons having travelled paths with larger $q$ (case b in Fig. 13b). For PS310, we expect from our qualitative arguments, that this does not alter the number of detectable MS photons very much, due to its strong forward scattering characteristic. Thus, for PnBAPS118, $q_{MS}$ will show a stronger increase than for PS310. This is experimentally supported by the approximately linear decrease of $N_{MS}$ in Fig. 13a.

A different way of increasing $q_{MS}$ is an increase of the detection volume. Experimentally this can be readily realized by removing the lens and the aperture. Then the detection $k_N$ and $q_{SS}$ are defined by the alignment of the fiber optics with $\Delta q_{SS}$ defined by the acceptance angle of the grin lens. The range of $\Delta q_{SS}$ about $q_{SS}$ is thus increased, and the diffusional broadening of the SS-signal gets larger. At the same time, a slightly conical detection volume of approximately doubled cross section is realized allowing the detection of light scattered of a slightly larger portion of the illumination cone produced by the forgoing scattering events. The probability of detecting photons having taken paths not close to the shortest possible ones increases. Therefore the average $q_{MS}$ will increase at constant $n$. For a suspension of PS310 at $n = 1.3 \times 10^{16}$ m$^{-3}$ we obtained an increase of $w_{MS}$ from 95 Hz to 135 Hz corresponding to an increase in $q_{MS}$ by a factor of 5. Also this observation strongly supports our qualitative arguments concerning the influence of detection volume restriction.

We therefore think that the moderate width of the MS-Lorentzian is caused by super-heterodyning, a strongly restricted detection volume, a bias towards selected (ranges of) scattering vectors in SS through P(q) and S(q) and a bias toward short photon paths given by attenuation. Furthermore, attenuation of the reference beam and along the scattering paths combined with detection volume restriction yields the much weaker than expected dependence of the detected integrated spectral power on $n$.